\pdfoutput=1
\RequirePackage{ifpdf}
\ifpdf 
\documentclass[pdftex]{sigma}
\else
\documentclass{sigma}
\fi

\numberwithin{equation}{section}

\begin{document}
\allowdisplaybreaks

\renewcommand{\PaperNumber}{019}

\FirstPageHeading

\ShortArticleName{Tippe Top Equations and Equations for the Related Mechanical Systems}

\ArticleName{Tippe Top Equations and Equations\\ for the Related Mechanical Systems}

\Author{Nils RUTSTAM}

\AuthorNameForHeading{N.~Rutstam}

\Address{Department of Mathematics, Link\"{o}ping University, Link\"{o}ping, Sweden}
\Email{\href{mailto:ergoroff@hotmail.com}{ergoroff@hotmail.com}}

\ArticleDates{Received October 21, 2011, in f\/inal form March 27, 2012; Published online April 05, 2012}

\Abstract{The equations of motion for the rolling and gliding Tippe Top (TT) are nonintegrable and dif\/f\/icult to analyze. The only existing arguments about TT inversion are based on analysis of stability of asymptotic solutions and the LaSalle type theorem. They do not explain the dynamics of inversion.
To approach this problem we review and analyze here the equations of motion for the rolling and gliding TT in three equivalent forms, each one providing dif\/ferent bits of information about motion of TT. They lead to the main equation for the TT, which describes well the oscillatory character of motion of the symmetry axis $\mathbf{\hat{3}}$ during the inversion.
We show also that the equations of motion of TT give rise to equations of motion for two other simpler mechanical systems: the gliding heavy symmetric top and the gliding eccentric cylinder. These systems can be of aid in understanding the dynamics of the inverting TT.}


\Classification{70F40; 74M10; 70E18; 70E40; 37B25}

The Tippe Top (TT) is known for its counterintuitive behaviour; when it is spun, its rotation axis turns upside down and its center of mass rises. Its behaviour is presently understood through analysis of stability of its straight and inverted spinning solution. The energy of TT is a monotonously decreasing function of time and it appears that it is a suitable Lyapunov function for showing instability of the straight spinning solution and stability of the inverted spinning solution \cite{Mars,Eben,Karap3,Karap2,RSG}. The energy is also a suitable LaSalle function to conclude that for suf\/f\/iciently large angular momentum $\mathbf{L}$ directed close to the vertical $\hat{z}$-direction the inverted spinning solution is asymptotically attracting~\cite{Mars,RSG}.

Since the 1950s \cite{Hugen} it is also known how TT has to be built for the inversion to take place, more precisely that if $0<\alpha<1$ denotes the eccentricity of the center of mass then the quotient of moments of inertia $\gamma=\frac{I_1}{I_3}$ has to satisfy the inequality $1-\alpha<\gamma<1+\alpha$.
It is also understood that the gliding friction is necessary for converting rotational energy to potential energy.

 When it comes to describing the actual dynamics of TT, there is very little known. In several papers \cite{Coh,Eben,Or,Ued} numerical results of how the Euler angles $(\theta(t),\varphi(t),\psi(t))$ change during inversion are presented, but an understanding of qualitative features of solutions (as well as details of physical forces and torques acting during the inversion) remains a rather unexplored f\/ield.

There is also an alternative approach \cite{Rau,Nisse} through the main equation of the TT that shows the source of oscillatory behaviour of $\theta(t)$ and of the remaining variables $\varphi(t)$, $\psi(t)$. The main dif\/f\/iculty in making further progress lies in the complexity of the TTs equations of motion.

 The TT is described, in the simplest setting, by six nonlinear dynamical equations for six variables $(\theta(t),\dot{\theta}(t),\dot{\varphi}(t),\omega_3(t),\nu_x(t),\nu_y(t))$. Analysis is slightly simplif\/ied by the fact that these equations admit one integral of motion, the Jellett integral, and that the energy function is monotonously decreasing in time.
To make further progress in reading of\/f the dynamical content of the equations it is useful to study equations in all possible forms, to study special solutions and certain limiting cases of TT equations.
By special cases we mean reduced forms of TT equations obtained by imposing time-preserved constraints and some limiting forms of TT equations. For instance in this paper, we show that TT equations reduce, in a suitable limit, to the equations for a gliding heavy symmetric top. These reduced and limiting equations are usually simpler to study and the lessons learned from these cases are useful for better understanding of dynamics of the full TT equations.

 In this paper we review the results for the TT, modelled as an axisymmetric sphere, with the aim of understanding how a rigorous statement about dynamics of TT can be derived from full equations for TT and the role of extra assumptions made in earlier works.

\section{Notation and a model for the TT}

Inversion of the TT can be divided into four signif\/icant phases. During the f\/irst phase the TT initially spins with the handle pointing vertically up, but then starts to wobble. During the second phase the wobbling leads to inversion, in which the TT f\/lips so that its handle is pointing downwards. In the third phase the inversion is completed so the TT is spinning upside down. In the last phase it is stably spinning on its handle until the energy dissipation, due to the spinning friction, makes it fall down. Demonstrations shows that the lifespan of the f\/irst three phases are usually short, with the middle inversion phase being the shortest. The fourth phase lasts signif\/icantly longer.

 This empirical description of the inversion behaviour of the TT is comprehensible for anyone who observes the TT closely, but it lacks precision and clarity when it comes to answering the question of how this inversion occurs, and under which circumstances.

\begin{figure}[ht]
\centering
\includegraphics[scale=0.90]{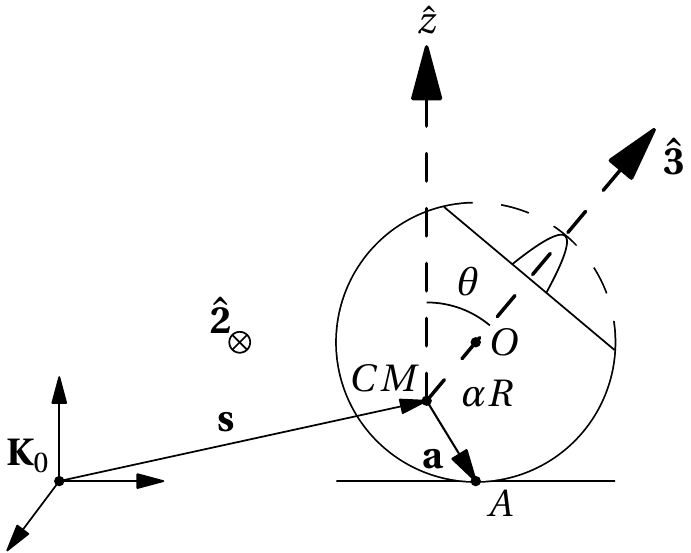}
\caption{Diagram of the TT model. Note that $\mathbf{a}=R\alpha\mathbf{\hat{3}}-R\hat{z}$.\label{TT_diagram}}
\end{figure}

  We model the TT as a sphere with radius $R$ and axially symmetric distributed mass $m$. It is in instantaneous contact with the supporting plane at the point $A$. The center of mass $CM$ is shifted from the geometric center~$O$ along the symmetry axis by $\alpha R$, where $0<\alpha<1$ (see Fig.~\ref{TT_diagram}). This model has been used in many works, for example by Hugenholtz~\cite{Hugen}, Cohen~\cite{Coh} and Karapetyan~\cite{Karap2,Karap1}. An alternative model of the TT could be to consider the TT as two spherical segments joined by a rigid rod, as proposed in~\cite{Karap4}.

 We choose a f\/ixed inertial reference frame $\mathbf{K}_0=(\widehat{X},\widehat{Y},\widehat{Z})$ with $\widehat{X}$ and $\widehat{Y}$ parallel to the supporting plane and with vertical $\widehat{Z}$. Let $\mathbf{K}=(\hat{x},\hat{y},\hat{z})$ be a frame def\/ined through rotation around $\widehat{Z}$ by an angle $\varphi$, where $\varphi$ is the angle between the plane spanned by $\widehat{X}$ and $\widehat{Z}$ and the plane spanned by the points $CM$, $O$ and $A$. The third reference frame is $\mathbf{\tilde{K}}=(\mathbf{\hat{1}},\mathbf{\hat{2}},\mathbf{\hat{3}})$, with origin at~$CM$, def\/ined through a rotation around $\hat{y}$ by an angle $\theta$, where $\theta$ is the angle between~$\widehat{Z}$ and the symmetry axis. The frame $\mathbf{\tilde{K}}$ is not fully f\/ixed in body since it rotates. We let $\mathbf{\hat{3}}$ be parallel to the symmetry axis.

 The Euler angles of the body relative to $\mathbf{K}_0$ are $(\theta,\varphi,\psi)$. The reference frame $\mathbf{K}$ rotates with the angular velocity $\dot{\varphi}\hat{z}$ w.r.t.\ $\mathbf{K}_0$ and the angular velocity of the frame $\mathbf{\tilde{K}}$ w.r.t.~$\mathbf{K}_0$ is
 \[
 \boldsymbol{\omega}_{\textrm{ref}}=\dot{\theta}\mathbf{\hat{2}}+\dot{\varphi}\hat{z}
 =-\dot{\varphi}\sin\theta\mathbf{\hat{1}}+\dot{\theta}\mathbf{\hat{2}}+\dot{\varphi}\cos\theta\mathbf{\hat{3}}.
 \]
The total angular velocity of the TT is found by adding the rotation around the symmetry axis~$\mathbf{\hat{3}}$:
\[
\boldsymbol{\omega}=\boldsymbol{\omega}_{\textrm{ref}}+\dot{\psi}\mathbf{\hat{3}}
=-\dot{\varphi}\sin\theta\mathbf{\hat{1}}+\dot{\theta}\mathbf{\hat{2}}+(\dot{\psi}+\dot{\varphi}\cos\theta)\mathbf{\hat{3}},
\]
and we shall refer to the third component of this vector as $\omega_3=\dot{\psi}+\dot{\varphi}\cos\theta$.

 For an arbitrary vector $\mathbf{B}$ in the rotating frame $\mathbf{\tilde{K}}$ the time-derivative is
 \[
 \mathbf{\dot{B}}=\frac{d}{dt}\left(B_1\mathbf{\hat{1}}+B_2\mathbf{\hat{2}}+B_3\mathbf{\hat{3}}\right)=\left(\frac{\partial\mathbf{B}}{\partial t}\right)_{\mathbf{\tilde{K}}}+\boldsymbol{\omega}_{\textrm{ref}}\times\mathbf{B}.
 \]
Here the f\/irst term $\dot{B}_1\mathbf{\hat{1}}+\dot{B}_2\mathbf{\hat{2}}+\dot{B}_3\mathbf{\hat{3}}$ is the time-derivative of each component describing the change of the vector $\mathbf{B}$ w.r.t.\ the frame $\mathbf{\tilde{K}}$ and $\boldsymbol{\omega}_{\textrm{ref}}\times\mathbf{B}$ the change of $\mathbf{B}$ due to rotation of $\mathbf{\tilde{K}}$ w.r.t.\ the frame $\mathbf{K}$.

 Let $\mathbf{a}=R(\alpha\mathbf{\hat{3}}-\hat{z})$ be the vector from $CM$ to the point of contact $A$.
 Newton's equations for the TT describe the motion of the $CM$ and rotation around the $CM$:
\begin{gather}
m\mathbf{\ddot{s}}=\mathbf{F}-mg\hat{z},\qquad
\mathbf{\dot{L}}=\mathbf{a}\times\mathbf{F}.
\label{EOM_TT_2}
\end{gather}
The external friction-reaction force $\mathbf{F}$ acts at the point of support $A$. The angular momentum is $\mathbf{L}=\mathbb{I}\boldsymbol{\omega}$, where $\mathbb{I}$ is the inertia tensor w.r.t.\ $CM$ and $\boldsymbol{\omega}$ is the angular velocity. If we denote the principal moments of inertia by $I_1$, $I_2$ and $I_3$ and note that $I_1=I_2$ due to the axial symmetry, the inertia tensor has the form $\mathbb{I}=I_1(\mathbf{\hat{1}}\mathbf{\hat{1}}^t+\mathbf{\hat{2}}\mathbf{\hat{2}}^t)+I_3\mathbf{\hat{3}}\mathbf{\hat{3}}^t$ (the superscript $t$ meaning here the transpose). The motion of the symmetry axis $\mathbf{\hat{3}}$ is described by the kinematic equation $\mathbf{\dot{\mathbf{\hat{3}}}}=\boldsymbol{\omega}\times\mathbf{\hat{3}}=\frac{1}{I_1}\mathbf{L}\times\mathbf{\hat{3}}.$

 We shall in our model of the TT assume that the point $A$ is always in contact with the plane, which can be expressed as the \emph{contact criterion} $\hat{z}\cdot(\mathbf{s}(t)+\mathbf{a}(t))\stackrel{t}{\equiv}0$. Since this is an identity with respect to time $t$ all its time derivatives have to vanish as well; in particular, $\hat{z}\cdot(\mathbf{\dot{s}}+\boldsymbol{\omega}\times\mathbf{a})=0$. So the velocity $\mathbf{v}_A(t)=\mathbf{\dot{s}}+\boldsymbol{\omega}\times\mathbf{a}$, which is the velocity of the point in the TT that is in instantaneous contact with the plane of support at time $t$, will also have zero $\hat{z}$-component.

 For the force acting at the point $A$ we assume $\mathbf{F}=g_n\hat{z}-\mu g_n\mathbf{v}_A$. This force consists of a normal reaction force and of a friction force $\mathbf{F}_f$ of viscous type, where $g_n$ and $\mu$ are non-negative. Other frictional forces due to spinning and rolling will be ignored. Since we have one scalar constraint equation $\hat{z}\cdot(\mathbf{s}+\mathbf{a})=0$, the component describing the vertical reaction force~$g_n\hat{z}$ is dynamically determined from the second time-derivative of the constraint. The planar component of $\mathbf{F}$ has to be specif\/ied independently to make the equations~\eqref{EOM_TT_2} fully determined. In our model we take $\mathbf{F}_f=-\mu g_n\mathbf{v}_A=-\mu(\mathbf{L},\mathbf{\hat{3}},\mathbf{\dot{s}},\mathbf{s},t)g_n(t)\mathbf{v}_A$ as the planar component.

This way of writing the friction force indicates that it acts against the gliding velocity~$\mathbf{v}_A$ and that the friction coef\/f\/icient can in principle depend on all dynamical variables and on time~$t$. We keep here also the factor~$g_n(t)$ to indicate that the friction is proportional to the value of the reaction force and that we are interested only in such motions where $g_n(t)\geq 0$.

 These assumptions about the external force are common in most of the literature about the TT. Cohen~\cite{Coh} states that a frictional force opposing the motion of the contact point is the only external force to act in the supporting plane. He used this assumption in a  numerical simulation that demonstrates that inversion of TT occurs. This showed the ef\/fectiveness of the model and in later works such as~\cite{Cio,Eben,Moff,Moff2,RSG}, the external force def\/ined in this way is used without comment. Or \cite{Or} discusses an inclusion of a nonlinear Coulomb-type friction in the external force along with the viscous friction. A~Coulomb term would in our notation look like $-\mu_C g_n\mathbf{v}_A/|\mathbf{v}_A|$, where~$\mu_C$ is a coef\/f\/icient. Numerical simulation shows that this Coulomb term can contribute to inversion, but has weaker ef\/fect. Bou-Rabee et al.~\cite{Mars} use this result to argue for only using an external force with a~normal reaction force and a frictional force of viscous type in the model of the TT, since the nonlinear Coulomb friction only results in algebraic destabilization of the initially spinning TT, whereas the viscous friction gives exponential destabilization.

 The above model of the external force becomes signif\/icant if we consider the energy function for the rolling and gliding TT
\begin{gather*}
E=\frac{1}{2}m\mathbf{\dot{s}}^2+\frac{1}{2}\boldsymbol{\omega}\cdot\mathbf{L}+mg\mathbf{s}\cdot\hat{z}.
\end{gather*}
When the energy is dif\/ferentiated the equations of motion yield $\dot{E}=\mathbf{F}\cdot\mathbf{v}_A$ and if the reaction force is $\mathbf{F}=g_n\hat{z}-\mu g_n\mathbf{v}_A$, then $\dot{E}=-\mu g_n\mathbf{v}_{A}^2$.
Thus for this model of a rolling and gliding TT, the energy is decreasing monotonically, as expected for a dissipative system.

 During the inversion of the TT the $CM$ is lifted up by  $2R\alpha$, which increases the potential energy by $2mgR\alpha$. This increase can only happen at the expense of the kinetic energy $T=\frac{1}{2}m\mathbf{\dot{s}}^2+\frac{1}{2}\boldsymbol{\omega}\cdot\mathbf{L}$ of the TT.
Analysis of the inversion must address how this transfer of energy occurs in the context of the friction model. The following proposition, due to an argument made by Del~Campo \cite{DelCampo}, gives an idea of how it works.

\begin{proposition}\label{DelCampo}
The component of the gliding friction that is perpendicular to the $(\hat{z},\mathbf{\hat{3}})$-plane is the only force enabling inversion of TT.
\end{proposition}
\begin{proof}
Inversion starts with the TT spinning in an almost upright position and ends with the TT spinning upside down, so the process of inversion requires transfer of energy from the kinetic term $\frac{1}{2}\boldsymbol{\omega}\cdot\mathbf{L}$ to the potential term $mg\mathbf{s}\cdot\hat{z}$. The angular momentum at the initial position~$\mathbf{L}_0$ and the f\/inal position $\mathbf{L}_1$ are both almost vertical, and the value has been reduced: $|\mathbf{L}_1|<|\mathbf{L}_0|$.

We let the reaction force $\mathbf{F}$ be split into $\mathbf{F}_R+\mathbf{F}_{f||}+\mathbf{F}_{f\perp}$, where $\mathbf{F}_R=g_n\hat{z}$ is the vertical reaction force, $\mathbf{F}_{f||}$ is the component of the planar friction force parallel to the $(\hat{z},\mathbf{\hat{3}})$-plane, and~$\mathbf{F}_{f\perp}$ is the component of the planar friction force that is perpendicular to this plane. We thus have
\begin{gather*}
\mathbf{\dot{L}}=\mathbf{a}\times(\mathbf{F}_R+\mathbf{F}_{f||}+\mathbf{F}_{f\perp}),
\end{gather*}
which implies that
\begin{gather*}
\mathbf{\dot{L}}\cdot\hat{z}=\left[\mathbf{a}\times(\mathbf{F}_R+\mathbf{F}_{f||})\right]
\cdot\hat{z}+(\mathbf{a}\times\mathbf{F}_{f\perp})\cdot\hat{z}=(\mathbf{a}\times\mathbf{F}_{f\perp})\cdot\hat{z}<0.
\end{gather*}
We see that without the friction force $\mathbf{F}_{f\perp}$ there is no reduction of $\mathbf{L}\cdot\hat{z}$ and no transfer of energy to the potential term. The torque $\mathbf{a}\times(\mathbf{F}_R+\mathbf{F}_{f||})$ is parallel to the plane of support and causes only precession of the vector~$\mathbf{L}$.
\end{proof}

\begin{figure}[ht]\centering
\includegraphics[scale=0.65]{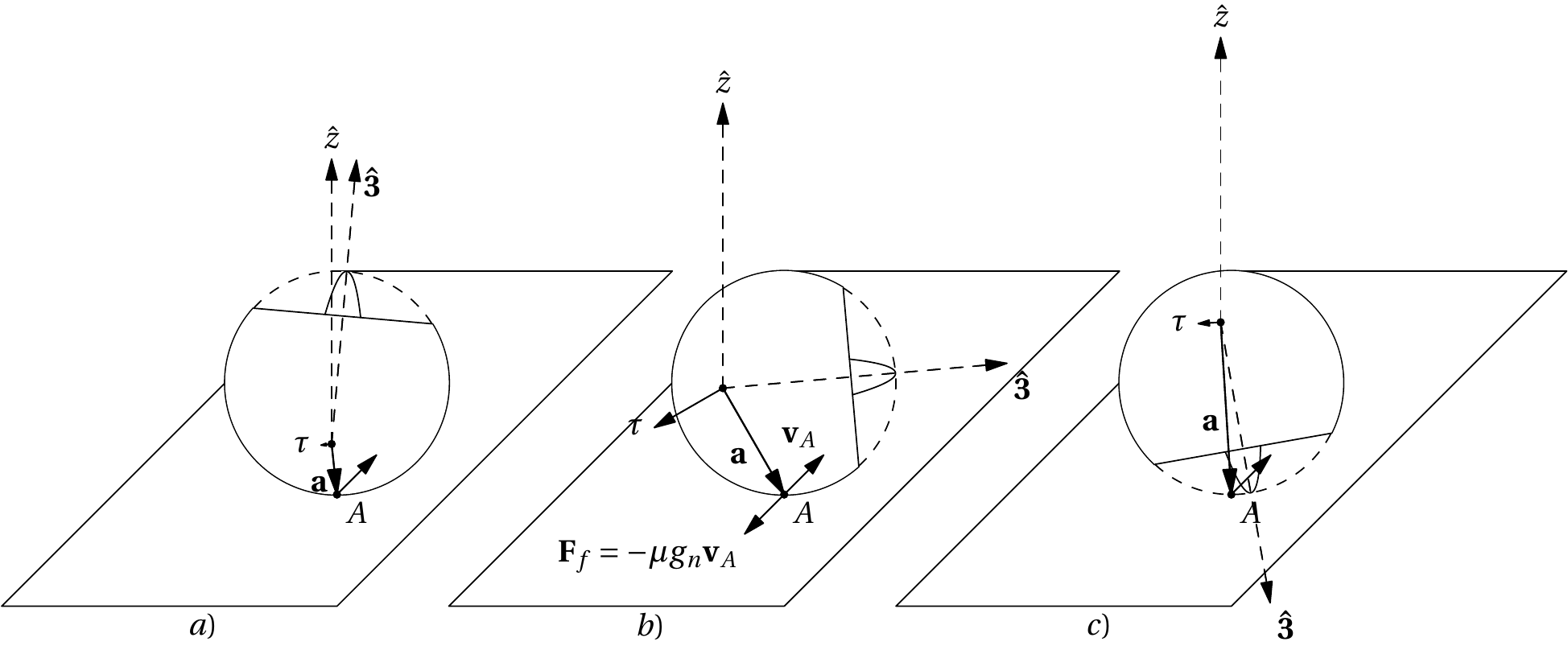}
\caption{The TT at dif\/ferent inclination angles during inversion.\label{Turning_TT}}
\end{figure}

Fig.~\ref{Turning_TT} illustrates this. The rotational gliding of the TT at the contact point $A$ creates an opposing force, which gives rise to an external torque $\boldsymbol{\tau}$ that reduces the $\hat{z}$-component of the angular momentum $\mathbf{L}$ and transfers energy from rotational kinetic energy into the potential energy. As a consequence, the $CM$ of the TT rises.

Indeed, in Fig.~\ref{Turning_TT}a we see that $\mathbf{L}$ is closely aligned with the $\hat{z}$-axis so the angular velocity is pointing almost upwards and the contact velocity~$\mathbf{v}_A$ is pointing into the plane of the picture. In Fig.~\ref{Turning_TT}b it is clear that $\mathbf{a}\times\mathbf{F}_f=\mathbf{a}\times(-\mu g_n\mathbf{v}_A)$ gives a torque $\boldsymbol{\tau}$ that has a negative $\hat{z}$-component. It is $\hat{z}\cdot(\mathbf{a}\times(-\mu g_n\mathbf{v}_A))=-(\mu g_n R\alpha\sin\theta)\mathbf{v}_{A}\cdot\mathbf{\hat{2}}$. This component has larger magnitude in Fig.~\ref{Turning_TT}b when the inclination angle is closer to $\pi/2$ and then decreases when the TT approaches the inverted position (Fig.~\ref{Turning_TT}c).

This may help us to understand the observed phenomenon that the inversion of TT is fast during the middle phase but is slower in the initial and in the f\/inal phase, since the $\hat{z}$-component of the torque responsible for transfer of energy and inversion is larger in the middle phase.

\subsection{The vector and the Euler angle forms of equations of TT}

We return to the equations of motion for the TT in vector form
\begin{gather}
m\mathbf{\ddot{s}}=\mathbf{F}-mg\hat{z},\qquad
\mathbf{\dot{L}}=\mathbf{a}\times\mathbf{F},\qquad
\mathbf{\dot{\mathbf{\hat{3}}}}=\frac{1}{I_1}\mathbf{L}\times\mathbf{\hat{3}}.\label{EOM_TT_3}
\end{gather}
The contact constraint $\hat{z}\cdot(\mathbf{s+a})\stackrel{t}{\equiv}0$ is a scalar equation which further reduces this system. As previously mentioned, all time derivatives of the contact constraint also have to vanish identically. The f\/irst derivative says that the contact velocity has to be in the plane at all times:
\[
\hat{z}\cdot(\mathbf{\dot{s}}+\boldsymbol{\omega}\times\mathbf{a})=\hat{z}\cdot\mathbf{v}_A=0.
 \]
 The second derivative gives that the contact acceleration is also restricted to the plane:
 \[
 \frac{d^2}{dt^2}\hat{z}\cdot(\mathbf{s+a})=\hat{z}\cdot\mathbf{\dot{v}}_A=0.
 \]
 The vertical component of~$\mathbf{F}$ can be determined from
 \[
 \hat{z}\cdot\mathbf{F}=\hat{z}\cdot\left(m\mathbf{\ddot{s}}+mg\hat{z}\right)=-m\frac{d}{dt}(\boldsymbol{\omega}\times\mathbf{a})\cdot\hat{z}+mg.
 \]

The planar components of $\mathbf{F}$ have to be def\/ined to specify the equations of motion, and this makes it impossible to distinguish between the friction force and the planar components of the reaction force. This is why we need to specify the planar part of the reaction force in our model to make equations~\eqref{EOM_TT_3} complete.

The specif\/ics of the functions in the gliding friction, $\mu(\mathbf{L},\mathbf{s},\mathbf{\dot{s}},\mathbf{\hat{3}},t)$ and $g_n(\mathbf{L},\mathbf{s},\mathbf{\dot{s}},\mathbf{\hat{3}},t)$, other than that they are greater than zero are not relevant in the asymptotic analysis, but we can get the value of $g_n$ by taking the second derivative of the contact constraint and using the equations of motion:
\[
0=\hat{z}\cdot\left(\mathbf{\ddot{s}}+\frac{d}{dt}(\boldsymbol{\omega}\times\mathbf{a})\right)
=\frac{1}{m}(g_n-mg)+\frac{R\alpha}{I_1}\hat{z}\cdot\big(\mathbf{\dot{L}}\times\mathbf{\hat{3}}+\mathbf{L}\times\mathbf{\dot{\mathbf{\hat{3}}}}\big),
\]
so that
\begin{gather*}
g_n=\frac{mgI_{1}^2+mR\alpha(\mathbf{\hat{3}}_{\hat{z}}L^2-L_{\hat{z}}L_{\mathbf{\hat{3}}})}{I_{1}^2
+mI_{1}R^2\alpha^2(1-\mathbf{\hat{3}}_{\hat{z}}^2)+mI_{1}R^2\alpha(\alpha\mathbf{\hat{3}}_{\hat{z}}-1)\mu\mathbf{v}_{A}\cdot\mathbf{\hat{3}}},
\end{gather*}
where $L_{\mathbf{\hat{3}}}=\mathbf{L}\cdot\mathbf{\hat{3}}$,
$L_{\hat{z}}=\mathbf{L}\cdot\hat{z}$, $\mathbf{\hat{3}}_{\hat{z}}=\mathbf{\hat{3}}\cdot\hat{z}$ and $L^2=|\mathbf{L}|^2$.

With $g_n$ def\/ined this way, we see that no further derivatives of the contact constraint are needed since the second derivative $\hat{z}\cdot(\mathbf{\ddot{s}}(t)+\frac{d}{dt}(\boldsymbol{\omega}\times\mathbf{a}))=0$ becomes an identity with respect to time after substituting solutions for $\mathbf{s}(t)$, $\boldsymbol{\omega}(t)$ and $\mathbf{a}(t)$. The contact constraint determines the vertical component of $\mathbf{s}$: $s_{\hat{z}}=-\mathbf{a}\cdot\hat{z}=-R(\alpha\mathbf{\hat{3}}_{\hat{z}}-1)$, and its derivative $\mathbf{\dot{s}}\cdot\hat{z}=-(\boldsymbol{\omega}\times\mathbf{a})\cdot\hat{z}$. Thus the original system can be written as
\begin{gather}
m\mathbf{\ddot{r}}=-\mu g_n\mathbf{v}_A,\qquad
\mathbf{\dot{L}}=\mathbf{a}\times\mathbf{F},\qquad
\mathbf{\dot{\mathbf{\hat{3}}}}=\frac{1}{I_1}\mathbf{L}\times\mathbf{\hat{3}}, \label{EOM_TT_constraint}
\end{gather}
where $\mathbf{r}=\mathbf{s}-s_{\hat{z}}\hat{z}$. This system has 10 unknown variables
$(\mathbf{L},\mathbf{r},\mathbf{\dot{r}},\mathbf{\hat{3}})$. Further, if we assume
$\mu=\mu(\mathbf{L},\mathbf{\dot{s}},\mathbf{\hat{3}})$ then this system does
not depend explicitly on $\mathbf{r}$, so we ef\/fectively have a system of 8 ODEs with 8 unknowns.

 A rolling and gliding rigid body with a spherical shape such as the TT admits one integral of motion, the \emph{Jellett integral}
\[ 
 \lambda=-\mathbf{L}\cdot\mathbf{a}=R(L_{\hat{z}}-\alpha L_{\mathbf{\hat{3}}}).
 \]
It is an integral since
\[
\dot{\lambda}=-\mathbf{\dot{L}}\cdot\mathbf{a}-\frac{d}{dt}(R\alpha\mathbf{\hat{3}}-R\hat{z})\cdot\mathbf{L}
=-(\mathbf{a}\times\mathbf{F})\cdot\mathbf{a}-\mathbf{L}\cdot\left(\frac{R\alpha}{I_1}(\mathbf{L}\times\mathbf{\hat{3}})\right)=0.
\]

 We can rewrite the reduced equations of motion for the TT with the reaction force in explicit form
\begin{gather}
\frac{d}{dt}\left(\mathbb{I}\boldsymbol{\omega}\right)=\mathbf{a}\times\left(g_n\hat{z}-\mu g_n\mathbf{v}_A\right),\qquad
m\mathbf{\ddot{r}}=-\mu g_n\mathbf{v}_A, \label{EOM_TT_red1}
\end{gather}
using the Euler angle notation. The angular velocity has the form $\boldsymbol{\omega}=-\dot{\varphi}\sin\theta\mathbf{\hat{1}}+\dot{\theta}\mathbf{\hat{2}}+\omega_3\mathbf{\hat{3}}$.
We write the gliding velocity of the point of support as $\mathbf{v}_{A}=\nu_x\cos\theta\mathbf{\hat{1}}+\nu_y \mathbf{\hat{2}}+\nu_x\sin\theta\mathbf{\hat{3}}$, where $\nu_x$,~$\nu_y$ are components in the $\mathbf{\hat{2}}\times\hat{z}$ and $\mathbf{\hat{2}}$ direction (note here that $\hat{z}\cdot\mathbf{v}_A=0$ as expected). The equations~\eqref{EOM_TT_red1} have the following form in Euler angles:
\begin{gather}
 \label{EOM_TT_Euler_1}
 -I_1\ddot{\varphi}\sin\theta-2I_1\dot{\varphi}\dot{\theta}\cos\theta+I_3\omega_3\dot{\theta}=R(\alpha-\cos\theta)\mu g_n\nu_y,\\
 \label{EOM_TT_Euler_2}
 I_1\ddot{\theta}-I_1\dot{\varphi}^2\sin\theta\cos\theta+I_3\omega_3\dot{\varphi}\sin\theta=-R\alpha g_n\sin\theta+R\mu g_n\nu_x(1-\alpha\cos\theta),\\
 \label{EOM_TT_Euler_3}
 I_3\dot{\omega}_3=-R\mu g_n\nu_y\sin\theta,\\
 \label{EOM_TT_Euler_4}
 m\big(\dot{\nu}_x-\dot{\varphi}\nu_y+R(\ddot{\theta}(1-\alpha\cos\theta)+\alpha\dot{\theta}^2
 \sin\theta+\dot{\varphi}\sin\theta(\dot{\varphi}(\alpha-\cos\theta)+\omega_3))\big)=-\mu g_n\nu_x,\\
 \label{EOM_TT_Euler_5}
 m\big(\dot{\nu}_y+\dot{\varphi}\nu_x-R(\sin\theta(\ddot{\varphi}(\alpha-\cos\theta)
 +\dot{\omega}_3)+\dot{\theta}\cos\theta(2\dot{\varphi}(\alpha-\cos\theta)+\omega_3))\big)=-\mu g_n\nu_y.
\end{gather}
By solving this linear system for the highest derivative of each variable  $(\theta,\varphi,\omega_3,\nu_x,\nu_y)$, we obtain the system
\begin{gather}
\label{ddth}
\ddot{\theta}= \frac{\sin\theta}{I_1}\left(I_1\dot{\varphi}^2\cos\theta-I_3\omega_3\dot{\varphi}-R\alpha g_n\right)+\frac{R\mu g_n\nu_x}{I_1}(1-\alpha\cos\theta),\\
\label{ddph}\ddot{\varphi}= \frac{I_3\dot{\theta}\omega_3-2I_1\dot{\theta}\dot{\varphi}\cos\theta-\mu g_n\nu_y R(\alpha-\cos\theta)}{I_1\sin\theta},\\
\label{dot_om}\dot{\omega}_3= -\frac{\mu g_n\nu_y R\sin\theta}{I_3},\\
\label{nu_x}\dot{\nu}_x= \frac{R\sin\theta}{I_1}\big(\dot{\varphi}\omega_3\left(I_3(1-\alpha\cos\theta)-I_1\right)
+g_nR\alpha(1-\alpha\cos\theta)-I_1\alpha(\dot{\theta}^2+\dot{\varphi}^2\sin^2\theta)\big)\nonumber\\
\phantom{\dot{\nu}_x=}{}
 -\frac{\mu g_n\nu_x}{mI_1}\big( I_1+mR^2(1-\alpha\cos\theta)^2\big)+\dot{\varphi}\nu_y,\\
\dot{\nu}_y= -\frac{\mu g_n\nu_y}{mI_1 I_3}\big(I_1I_3+mR^2 I_3(\alpha-\cos\theta)^2+mR^2I_1\sin^2\theta\big)\nonumber\\
\hphantom{\dot{\nu}_y=}{}  +\frac{\omega_3\dot{\theta} R}{I_1}\left(I_3(\alpha-\cos\theta)+I_1\cos\theta\right)-\dot{\varphi}\nu_x,\label{nu_y}
\end{gather}
with
\begin{gather}
\label{g_n_TT_euler}
g_n=\frac{mgI_1+mR\alpha\big(\cos\theta(I_1\dot{\varphi}^2\sin^2\theta+I_1\dot{\theta}^2)
-I_3\dot{\varphi}\omega_3\sin^2\theta\big)}{I_1+mR^2\alpha^2\sin^2\theta-mR^2\alpha\sin\theta(1-\alpha\cos\theta)\mu\nu_x}.
\end{gather}
Above we have equations in an explicit form, which (if we add the equation $\dot{\theta}=\frac{d}{dt}(\theta)$) can be written as $(\dot{\theta}, \ddot{\theta},\ddot{\varphi},\dot{\omega}_3,\dot{\nu}_x,\dot{\nu}_y)$$=f(\theta,\dot{\theta},\varphi,\dot{\varphi},\psi,\dot{\psi},\nu_x,\nu_y)$. We see directly that the right hand sides of the equations are independent of $\varphi$ and $\psi$, which shows that we only need to consider equations for~$\theta$, $\dot{\theta}$, $\dot{\varphi}$, $\omega_3$, $\nu_x$ and $\nu_y$. So ef\/fectively the system~\eqref{EOM_TT_constraint} has 6 unknowns. Further, since we have the Jellett integral $\lambda=-\mathbf{L\cdot a}=RI_1\dot{\varphi}\sin^2\theta-R(\alpha-\cos\theta)I_3 \omega_3$, the number of unknowns can be reduced to 5 on each surface of constant value of~$\lambda$.

\subsection{Equations in Euler angles with respect to the rotating reference system~$\mathbf{K}$}

The equations of motion derived above were formulated and calculated with respect to the $(\mathbf{\hat{1}},\mathbf{\hat{2}},\mathbf{\hat{3}})$-system. We can arrive at an equivalent set of equations by formulating them in the $(\hat{x},\hat{y},\hat{z})$-system. This is the approach used in \cite{Moff2} and \cite{Ued}. We f\/irst use the relations
\begin{gather*}
\hat{x}=\cos\theta\mathbf{\hat{1}}+\sin\theta\mathbf{\hat{3}} , \qquad
\hat{y}=\mathbf{\hat{2}} \qquad \mbox{and} \qquad \hat{z}=-\sin\theta\mathbf{\hat{1}}+\cos\theta\mathbf{\hat{3}}
\end{gather*}
to get the angular velocity
\[
\boldsymbol{\omega}=(\omega_3-\dot{\varphi}\cos\theta)\sin\theta\hat{x}+\dot{\theta}\hat{y}+(\dot{\varphi}+\cos\theta(\omega_3-\dot{\varphi}\cos\theta))\hat{z},
\]
 and
\[
\mathbf{L}=(I_3\omega_3-I_1\dot{\varphi}\cos\theta)\sin\theta\hat{x}+I_1\dot{\varphi}\hat{y}+(I_1\dot{\varphi}
+\cos\theta(I_3\omega_3-I_1\dot{\varphi}\cos\theta))\hat{z}.
\]

 We need also note that the frame $\mathbf{K}=(\hat{x},\hat{y},\hat{z})$ rotates with the angular velocity $\dot{\varphi}\hat{z}$. Thus starting from the equations of motion in the form~\eqref{EOM_TT_red1}
\begin{gather*}
\mathbf{\dot{L}}=\mathbf{a}\times\left(g_n\hat{z}-\mu g_n\mathbf{v}_A\right),\qquad
 m\mathbf{\ddot{r}}=-\mu g_n\mathbf{v}_A,
\end{gather*}
(where $\mathbf{r}=s_{\hat{x}}\hat{x}+s_{\hat{y}}\hat{y}$) and taking into account that the time-derivative is taken with respect to the rotating frame $\mathbf{K}$ (which means for an arbitrary vector~$\mathbf{B}$ in this frame that $\mathbf{\dot{B}}=\left(\frac{\partial\mathbf{B}}{\partial t}\right)_{\mathbf{K}}+\dot{\varphi}\hat{z}\times\mathbf{B}$), and that the gliding velocity in the $(\hat{x},\hat{y},\hat{z})$-system is $\mathbf{v}_A=\nu_x\hat{x}+\nu_y\hat{y}$. We obtain in the Euler angles the following equations:
\begin{gather*}
 \frac{d}{dt}\left(\sin\theta(I_3\omega_3-I_1\dot{\varphi}\cos\theta)\right)=-R(1-\alpha\sin\theta)\mu g_n\nu_y ,\\
 I_1\ddot{\theta}+\dot{\varphi}\sin\theta(I_3\omega_3-I_1\dot{\varphi}\cos\theta)=-R\alpha g_n\sin\theta+R(1-\alpha\cos\theta)\mu g_n\nu_x ,\\
 I_1\ddot{\varphi}+\frac{d}{dt}\left(\cos\theta(I_3\omega_3-I_1\dot{\varphi}\cos\theta)\right)=-R\alpha\mu g_n\nu_y\sin\theta ,\\
 m\ddot{s}_{\hat{x}}=m\dot{\varphi}\dot{s}_{\hat{y}}-\mu g_n\nu_x,\\
 m\ddot{s}_{\hat{y}}=-m\dot{\varphi}\dot{s}_{\hat{x}}-\mu g_n\nu_y,
\end{gather*}
The advantage of rederiving the equations of motion in this frame is that it becomes easier to use a version of the gyroscopic balance condition. This condition, as formulated by Mof\/fatt and Shimomura~\cite{Moff}, says that for a rapidly precessing axisymmetric rigid body we will have
\begin{gather*}
\xi:=I_3\omega_3-I_1\dot{\varphi}\cos\theta\approx0.
\end{gather*}
This approximation, while somewhat applicable to such rigid bodies as a spinning egg, fails for the TT initially as Ueda et al.\ have pointed out~\cite{Ued}. This is since the TT starts with initial angle $\theta(0)\approx 0$ and the spin velocity~$\dot{\psi}$ about the symmetry axis is large, while a spinning egg (axisymmetric spheroid) is thought to start lying on the side, which corresponds to initial angle $\theta(0)\approx\frac{\pi}{2}$ and small~$\dot{\psi}$. Experiments have suggested that the condition $\xi\approx 0$ is approximately satisf\/ied during some phase of the inversion, and if we use the new variable $\xi=I_3\omega_3-I_1\dot{\varphi}\cos\theta$ instead of $\dot{\varphi}$,
then the equations of motion take a simpler form:
\begin{gather*}
 \sin\theta\dot{\xi}+\dot{\theta}\xi\cos\theta=-R(1-\alpha\sin\theta)\mu g_n\nu_y,\\
 I_1\ddot{\theta}+\dot{\varphi}\xi\sin\theta=-R\alpha g_n\sin\theta+R(1-\alpha\cos\theta)\mu g_n\nu_x,\\
 I_1\ddot{\varphi}+\dot{\xi}\cos\theta-\dot{\theta}\xi\sin\theta=-R\alpha\mu g_n\nu_y\sin\theta,\\
 m\ddot{s}_{\hat{x}}=m\dot{\varphi}\dot{s}_{\hat{y}}-\mu g_n\nu_x,\\
 m\ddot{s}_{\hat{y}}=-m\dot{\varphi}\dot{s}_{\hat{x}}-\mu g_n\nu_y.
\end{gather*}
The challenge is to show that $\xi\approx 0$ is true in a certain phase of the inversion of TT. We may note that in the special case of $I_1=I_3$, then $\xi=I_3\dot{\psi}$, so then the gyroscopic balance condition says that the spin about the symmetry axis becomes small during inversion.

In \cite{Moff} it has been argued that if we assume the condition $\xi\approx 0$ for a spheroid, then it is pos\-sible to obtain a simple equation for $\dot{\theta}(t)$ that can be integrated. Experimental results (e.g.~\mbox{\cite{Coh,Ued}}) show that $\theta$ nutates as it increases, which means that~$\theta$ does not increase monotonously. Thus for the TT an interesting problem is if we can def\/ine a mean value for the angle $\theta(t)$, $\langle\theta(t)\rangle$, which increases monotonously for the initial conditions such that TT inverts.

An estimate in~\cite{Ued} claims that a suitably def\/ined mean value of $\theta(t)$ increases during some time-interval early in the inversion phase. This is claimed under such assumptions that the average of $\dot{\xi}\sin\theta$ may be neglected and the average of $-R(1-\alpha\cos\theta)\mu g_n\nu_{y}$ is negative. Numerical simulations shows that there exist choices for parameters and initial conditions so that these assumptions lead to behaviour of solutions that is in agreement with the results of simulations. After this phase the authors assume that the TT is in a phase where~$\xi$ is close to zero, so the equation for $\dot{\theta}(t)$ shows that~$\theta(t)$ increases on average during the rest of inversion.

\subsection{The main equation for the TT}

An alternative form of the TT equations is inspired by the integrability of an axially symmetric sphere, rolling (without gliding) in the plane~\cite{Chap,Chap2,Routh}. The point $A$ is always in contact with the plane, and an algebraic constraint characterizing this motion is that the velocity of the point of contact vanishes, i.e.\ $\mathbf{v}_A=\mathbf{\dot{s}}+\boldsymbol{\omega}\times\mathbf{a}=0$. Thus we have $\mathbf{\dot{s}}=-\boldsymbol{\omega}\times\mathbf{a}$, and $\mathbf{F}$ can be eliminated from~\eqref{EOM_TT_2}:
\begin{gather*}
\frac{d}{dt}\left(\mathbb{I}\boldsymbol{\omega}\right)=m\mathbf{a}\times\left(-\frac{d}{dt}(\boldsymbol{\omega}\times\mathbf{a})+g\hat{z}\right),
\qquad\mathbf{\dot{\mathbf{\hat{3}}}}=\boldsymbol{\omega}\times\mathbf{\hat{3}}.
\end{gather*}
Notice that these equations dif\/fer from~\eqref{EOM_TT_red1} where assumptions about the reaction force $\mathbf{F}=g_n\hat{z}-\mu g_n\mathbf{v}_A$ are necessary to make the system def\/ined, while here the whole force $\mathbf{F}=m\mathbf{\ddot{s}}=m\frac{d}{dt}(\mathbf{a}\times\boldsymbol{\omega})$ is dynamically determined and does not enter into the equations.
In terms of the Euler angles we obtain three equations of motion for $\ddot{\theta}$, $\ddot{\varphi}$ and $\dot{\omega}_3$:
\begin{gather}
\ddot{\theta}= \frac{\sin\theta}{I_1+mR^2((\alpha-\cos\theta)^2+\sin^2\theta)}\big[\dot{\varphi}^2(I_1\cos\theta-mR^2(\alpha-\cos\theta)(1-\alpha\cos\theta))\nonumber\\
\phantom{\ddot{\theta}=}{}
-\dot{\varphi}\omega_3(mR^2(1-\alpha\cos\theta)+I_3)-\dot{\theta}^2mR^2\alpha-mgR\alpha\big],\label{EOM_rTT_Euler_1}\\
\nonumber\\
\label{EOM_rTT_Euler_2}\ddot{\varphi}= \frac{\omega_3\dot{\theta}(I_{3}^2
+mR^2I_3(1-\alpha\cos\theta))}{\sin\theta(I_1I_3+mR^2I_1\sin^2\theta+mR^2I_3(\alpha-\cos\theta)^2)}-\frac{2\dot{\theta}\dot{\varphi}\cos\theta}{\sin\theta},\\
\label{EOM_rTT_Euler_3}\dot{\omega}_3= -\omega_3\dot{\theta}\sin\theta\left(\frac{mR^2I_1\cos\theta+mR^2I_3(\alpha-\cos\theta)}{I_1I_3+mR^2I_1\sin^2\theta+mR^2I_3(\alpha-\cos\theta)^2}\right).
\end{gather}
This system admits three integrals of motion; the Jellett integral
\[
\lambda=RI_1\dot{\varphi}\sin^2\theta-RI_3\omega_3(\alpha-\cos\theta),
\]
 the energy
\begin{gather}
E= \frac{1}{2}mR^2\big[(\alpha-\cos\theta)^2(\dot{\theta}^2+\dot{\varphi}^2\sin^2\theta)
+\sin^2\theta(\dot{\theta}^2+\omega_{3}^2+2\omega_3\dot{\varphi}(\alpha-\cos\theta))\big]\nonumber\\
\phantom{E=}{} +\frac{1}{2}\big[I_1\dot{\varphi}^2\sin^2\theta+I_1\dot{\theta}^2+I_3\omega_{3}^2\big]+mgR(1-\alpha\cos\theta),\label{roll_energy}
\end{gather}
and the \emph{Routh integral}~\cite{Routh}, which can be found by integrating~\eqref{EOM_rTT_Euler_3}:
\begin{gather*}
D=\omega_3\big[I_1 I_3+mR^2I_3(\alpha-\cos\theta)^2+mR^2I_1\sin^2\theta\big]^{1/2}:=I_3\omega_3\sqrt{d(\cos\theta)},
\end{gather*}
where $d(\cos\theta)=\gamma+\sigma(\alpha-\cos\theta)^2+\sigma\gamma(1-\cos^2\theta)$, $\sigma=\frac{mR^2}{I_3}$ and $\gamma=\frac{I_1}{I_3}$.
If we are given the expressions for $\lambda$, $D$ and $E$ above then the equations
\begin{gather*}
\dot{\lambda}=0,\qquad \dot{D}=0,\qquad \dot{E}=0,
\end{gather*}
are (obviously) equivalent to the equations of motion \eqref{EOM_rTT_Euler_1}--\eqref{EOM_rTT_Euler_3}.

Since the equations of motion \eqref{EOM_rTT_Euler_1}--\eqref{EOM_rTT_Euler_3} do not depend on $\varphi$ and $\psi$, they constitute ef\/fectively a system of three equations for four unknowns $(\theta,\dot{\theta},\dot{\varphi},\omega_3)$, where the equation for $\theta$ is of second order. This is a fourth order dynamical system and the three integrals of motion $(\lambda,D,E)$ reduce the system to one equation. This is done by expressing $\dot{\varphi}$ and $\omega_3$ as functions of $\lambda$, $D$ and $\theta$:
\begin{gather*}
\omega_3=\dfrac{D}{I_{3}\sqrt{d(\cos\theta)}},\qquad \dot{\varphi}=\dfrac{\lambda\sqrt{d(\cos\theta)}+R(\alpha-\cos\theta)D}{RI_1\sin^2\theta\sqrt{d(\cos\theta)}}.
\end{gather*}
and by eliminating $\dot{\varphi}$ and $\omega_3$ from the energy~\eqref{roll_energy}. We obtain a separable dif\/ferential equation in $\theta$:
\begin{gather}
\label{MErTT}
E=g(\cos\theta)\dot{\theta}^2+V(\cos\theta,D,\lambda),
\end{gather}
where $g(\cos\theta)=\frac{1}{2}I_3(\sigma((\alpha-\cos\theta)^2+1-\cos^2\theta)+\gamma)$,
\begin{gather*}
V(z=\cos\theta,D,\lambda)=  mgR(1-\alpha z)+\frac{1}{2I_3 d(z)}\bigg(\frac{(\lambda\sqrt{d(z)}+R(\alpha-z)D)^2(\sigma(\alpha-z)^2+\gamma)}{R^2\gamma^2(1-z^2)}\nonumber\\
\phantom{V(z=\cos\theta,D,\lambda)=}{}
+D^2(\sigma(1-z^2)+1)+\frac{2}{R\gamma}D\sigma(\alpha-z)(\lambda\sqrt{d(z)}+R(\alpha-z)D)\bigg),
\end{gather*}
with $d(z)=\gamma+\sigma(\alpha-z)^2+\sigma\gamma(1-z^2)>0$.

 The function $V(z,D,\lambda)$ is the ef\/fective potential in the energy expression. This potential is well def\/ined and convex for $\theta$ in the interval $(0,\pi)$. We have that $V(\cos\theta,D,\lambda)\to\infty$ as $\theta\to 0,\pi$ if $\frac{D}{\lambda}\neq\frac{-\sqrt{d(\pm 1)}}{R(\alpha\mp 1)}$. Between these values, $V$ has one minimum, and the solution $\theta(t)$ stays between the two turning angles $\theta_0$ and $\theta_1$ determined from the equation $E=V(\cos\theta,D,\lambda)$ (i.e.\ when $\dot{\theta}=0$).

The ef\/fective potential for a rolling axisymmetric sphere was derived as early as~\cite{Chap2}. In Karapetyan and Kuleshov~\cite{Karap1} the same potential is found as a special case of a general method for calculating the potential of a conservative system with $n$ degrees of freedom and $k<n$ integrals of motion.
The potential for the rolling TT in the above form was derived and analysed in~\cite{Glad}.

\begin{figure}[ht]
\centering
\includegraphics[scale=0.60]{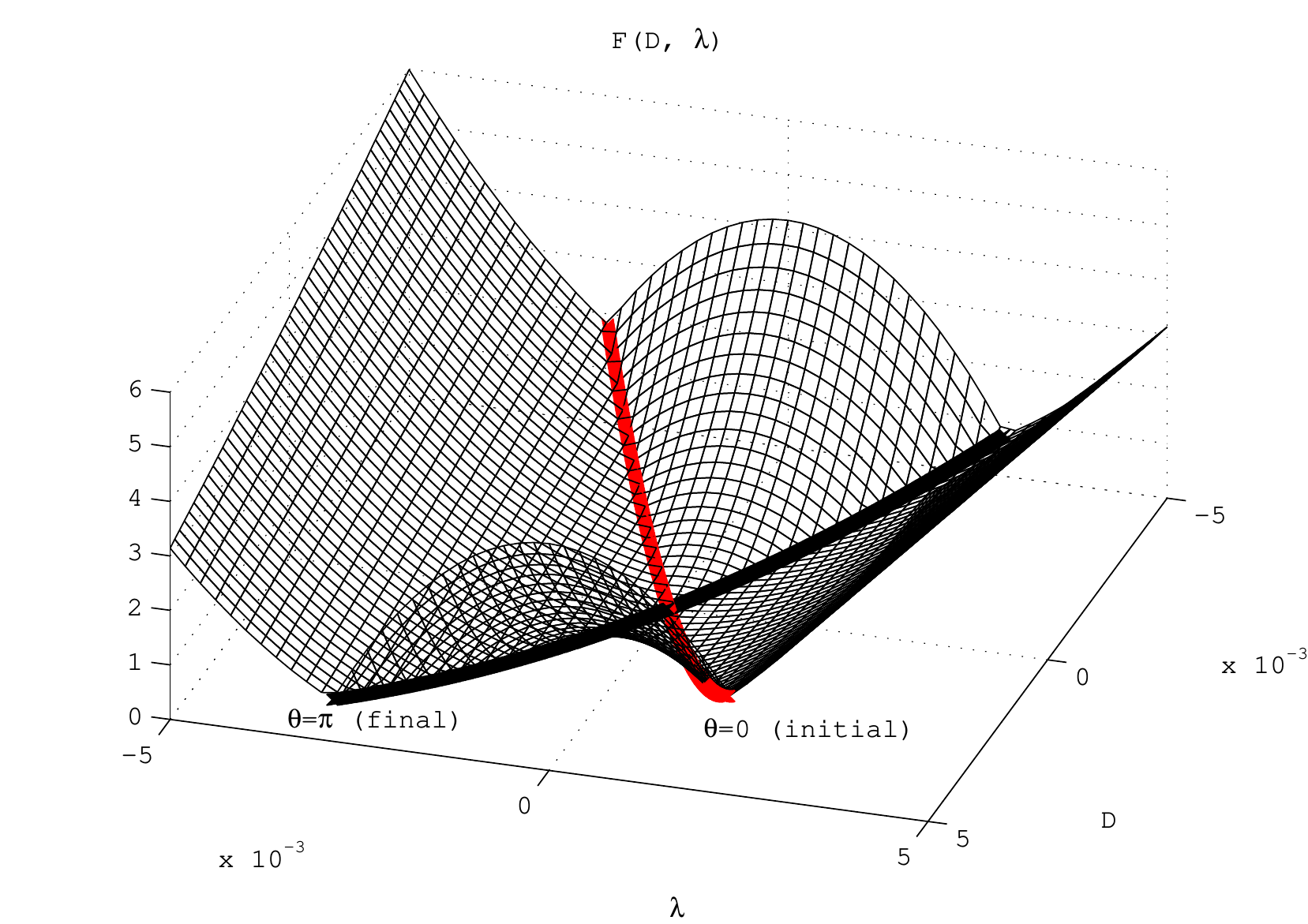}
\caption{Diagram for the minimal surface $F(D,\lambda)$.\label{TT_minimal}}
\end{figure}

 The dynamical states of the rolling TT, as described by the triple $(\lambda,D,E)$, are bounded below by the minimal value of the energy $E$, that is the surface given by $F(D,\lambda)=\!\!\min\limits_{z\in(-1,1)}\!\!V(z,D,\lambda)$. Varying the values for $\lambda$ and $D$, the minimal surface def\/ined by this equation can be generated. In Fig.~\ref{TT_minimal} we have taken the values of the parameters $m=0.02$, $R=0.02$, $\alpha=0.3$, $I_{1}=\frac{235}{625}mR^2$ and $I_3=\frac{2}{5}mR^2$. A similar surface showing the stationary points $\theta_{\min}(\lambda,D)$ for values of~$\lambda$ and~$D$ has also been discussed in \cite{Karap1}.
Points on and above this surface correspond to solutions to equation~\eqref{MErTT} for the rolling TT. In particular, points on the surface def\/ine precessional motions, and along the marked lines we have singled out the degenerate cases of precessional motion, the spinning motion where $\theta=0$ and $\theta=\pi$.

An inverting TT starts from a state close to a spinning, upright position and ends at a~state close to a spinning, inverted position. If we consider a trajectory giving the motion of an inverting TT, then in Fig.~\ref{TT_minimal} it has to stay in a vertical plane of $\lambda= {\rm const}$. It starts close above the line given by $\theta=0$ at the minimal surface and ends close above the line given by~$\theta=\pi$.

For better understanding of dynamics of the TT we can study time dependence of functions that are integrals for when the gliding velocity is zero. For very small $\mathbf{v}_A$ one may expect that these functions are changing slowly and are approximate integrals of motion.

 We already know that~$\lambda$ satisf\/ies~$\dot{\lambda}=0$ for a rolling and gliding TT.
For the total energy of TT, we know that $\dot{E}=\mathbf{F}\cdot\mathbf{v}_A\leq0$. However we shall consider the expression for the rolling energy of the TT
\begin{gather*}
\tilde{E}(\mathbf{L},\mathbf{\hat{3}},\mathbf{s})
=\frac{1}{2}m(\boldsymbol{\omega}\times\mathbf{a})^2+\frac{1}{2}\boldsymbol{\omega}\cdot\mathbf{L}+mg\mathbf{s}\cdot\hat{z}.
\end{gather*}
This is not the full energy, but the part not involving the quantity $\mathbf{v}_{A}$. Dif\/ferentiation yields:
\begin{gather}
\frac{d}{dt}\tilde{E}(\mathbf{L},\mathbf{\hat{3}},\mathbf{s}) =\frac{d}{dt}\left(E-\frac{1}{2}m\mathbf{v}_{A}\cdot\left(\mathbf{v}_A-2(\boldsymbol{\omega}\times\mathbf{a})\right)\right)\nonumber\\
 \label{diff_Et}
\phantom{\frac{d}{dt}\tilde{E}(\mathbf{L},\mathbf{\hat{3}},\mathbf{s})}{}
=\mathbf{F}\cdot\mathbf{v}_A-\mathbf{v}_{A}\cdot(m\mathbf{\ddot{s}})+m\mathbf{\dot{v}}_{A}\cdot(\boldsymbol{\omega}\times\mathbf{a})
=m\mathbf{\dot{v}}_{A}\cdot(\boldsymbol{\omega}\times\mathbf{a}).
\end{gather}
For the rolling and gliding TT the Routh function $D(\theta,\omega_3)=I_3\omega_3\sqrt{d(\cos\theta)}$, also changes in time:
\begin{gather}
\label{diff_Dt}
\frac{d}{dt}D(t)=\frac{\gamma mR\sin\theta}{\sqrt{d(\cos\theta)}}(\dot{\varphi}\nu_{x}+\dot{\nu}_y)=\frac{\gamma m}{\alpha\sqrt{d(\hat{z}\cdot\mathbf{\hat{3}})}}(\hat{z}\times\mathbf{a})\cdot\mathbf{\dot{v}}_A.
\end{gather}
We f\/ind from the expressions of $\lambda$ and $D(t)$ that
\begin{gather}
\label{TT_reduction}
\omega_3=\dfrac{D(t)}{I_3\sqrt{d(\cos\theta)}},\qquad \dot{\varphi}=\dfrac{\lambda\sqrt{d(\cos\theta)}+R(\alpha-\cos\theta)D(t)}{I_1\sqrt{d(\cos\theta)}\sin^2\theta},
\end{gather}
and eliminate $\dot{\varphi}$ and $\omega_3$ in the modif\/ied energy (same as equation~\eqref{roll_energy}) to get the Main Equation for the Tippe Top (METT):
\begin{gather}
\label{METT}
\tilde{E}(t)=g(\cos\theta)\dot{\theta}^2+V(\cos\theta,D(t),\lambda).
\end{gather}
It ostensibly has the same form as equation~\eqref{MErTT}, but now it depends explicitly on time through the functions $D(t)$ and $\tilde{E}(t)$. This means that separation of variables is not possible, but it is a~f\/irst order time dependent ODE easier to analyze than~\eqref{diff_Et} and~\eqref{diff_Dt}, provided that we have some quantitative knowledge about the functions~$D(t)$ and~$\tilde{E}(t)$.

 For generic values of $\lambda$ and $D$ the ef\/fective potential $V(\cos\theta,D(t),\lambda)$ has a single mini\-mum. Similarly, as in the pure rolling case, equation \eqref{METT} describes an oscillatory motion of the angle~$\theta(t)$ in the continuously deforming potential. For inverting solutions of TT, $\lambda=L_0 R(1-\alpha)=L_1 R(1+\alpha)$ (where $L_0$ and~$L_{1}$ are the values of $\mathbf{L}$ for $\theta=0$ and $\theta=\pi$, respectively), so $L_1=L_0\frac{1-\alpha}{1+\alpha}$ and $D$ changes from the value $D_0=L_0\sqrt{\gamma+\sigma(1-\alpha)^2}$ to $D_1=-L_0\frac{1-\alpha}{1+\alpha}\sqrt{\gamma+\sigma(1+\alpha)^2}$.

\begin{figure}[ht]
\centering
\includegraphics[scale=0.50]{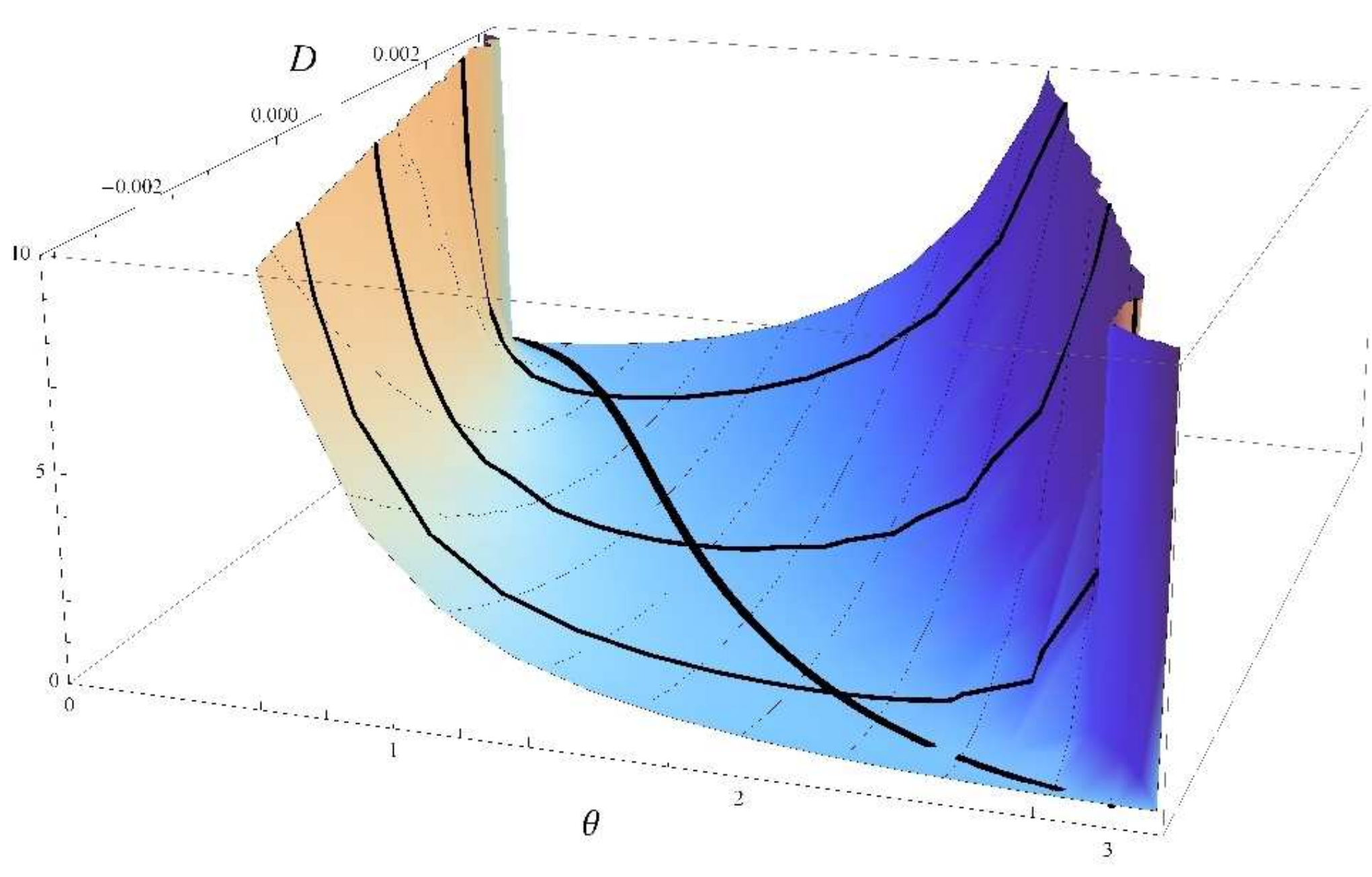}
\caption{Evolution of $V(\cos\theta,D(t),\lambda)$, $\theta\in(0,\pi)$, for $D$ between $D_0$ and $D_1$. The evolution of the minimum value of $V(\cos\theta,D(t),\lambda)$ is also marked. \label{Potential}}
\end{figure}
Fig.~\ref{Potential} shows how $V(\cos\theta,D(t),\lambda)$ deforms for $\theta\in(0,\pi)$, $D\in(D_1,D_0)$. We take here the same physical parameters $m=0.02$, $R=0.02$, $\alpha=0.3$, $I_{1}=\frac{235}{625}mR^2$ and $I_3=\frac{2}{5}mR^2$ as has been used for Fig.~\ref{TT_minimal}. We let the value of $\lambda$ be 10 times a threshold value (more on this in the next section). It is apparent here that the minimum value of the potential goes from being close to $\theta=0$ when $D$ is close to~$D_0$ and close to $\theta=\pi$ when $D$ is close to~$D_1$.

 The TT satisfying the contact criterion is still a system of 6 equations with 6 unknowns, but for these equations we have def\/ined three functions of time $\lambda$, $D(t)$ and $\tilde{E}(t)$, and have shown that the equations of motion~\eqref{EOM_TT_Euler_1}--\eqref{EOM_TT_Euler_5} are equivalent to the system
\begin{gather}
\frac{d}{dt}\lambda(\theta,\dot{\theta},\dot{\varphi},\omega_3)  =0,\nonumber\\
\frac{d}{dt}D(\theta,\omega_3) =\frac{\gamma m}{\alpha\sqrt{d(\hat{z}\cdot\mathbf{\hat{3}})}}(\hat{z}\times\mathbf{a})\cdot\mathbf{\dot{v}}_A=\frac{\gamma mR\sin\theta}{\sqrt{d(\cos\theta)}}(\dot{\varphi}\nu_{x}+\dot{\nu}_y),\nonumber\\
\frac{d}{dt}\tilde{E}(\theta,\dot{\theta},\dot{\varphi},\omega_3) =m(\boldsymbol{\omega}\times\mathbf{a})\cdot\mathbf{\dot{v}}_{A} \nonumber\\
\hphantom{\frac{d}{dt}\tilde{E}(\theta,\dot{\theta},\dot{\varphi},\omega_3)}{}
 = mR\big(\sin\theta(\dot{\varphi}(\alpha-\cos\theta)+\omega_3)(\dot{\varphi}\nu_x+\dot{\nu}_y)+ \dot{\theta}(1-\alpha\cos\theta)(\dot{\nu}_x-\dot{\varphi}\nu_y)\big),\nonumber\\
\frac{d}{dt}m\mathbf{\dot{r}} =-\mu g_n\mathbf{v}_A. \label{EOM_TT_diff}
\end{gather}
If we consider the motion of the TT as being determined by the three functions $(\lambda,D(t),\tilde{E}(t))$ satisfying these equations and connected by the METT, a useful method to investigate the inversion of TT crystallizes.

This system shows that the functions~$(D(t),\tilde{E}(t))$ allows us to analyze the equations of motion. If we have $D(t)$ and $\tilde{E}(t)$ given, then integrating the METT gives us~$\theta(t)$. With this information we can f\/ind the functions $\dot{\varphi}(t)$ and $\omega_{3}(t)$ from~\eqref{TT_reduction}. The equations containing derivatives of~$D(t)$ and~$\tilde{E}(t)$ in~\eqref{EOM_TT_diff} are two linear dif\/ferential equations for the velocities~$\nu_{x}(t)$ and~$\nu_y(t)$.

Thus the METT enables us to qualitatively study properties of a class of solutions that describes inverting solutions of the TT.

\section{Special solutions of TT equations}

We have already considered one class of special solutions, the pure rolling solutions satisfying the nongliding condition $\mathbf{v}_A=\mathbf{\dot{s}}+\boldsymbol{\omega}\times\mathbf{a}=0$. They are given by quadratures by solving the separable equation \eqref{MErTT}.

 A special subclass of rolling solutions to TT are solutions of the TT where an explicit assumption about the reaction force is taken ($\mathbf{F}=g_n\hat{z}-\mu g_n\mathbf{v}_A$). This means for rolling TT that we have now two conditions, $\mathbf{v}_A=0$ and $\mathbf{F}=g_{n}\hat{z}$. Note that the dynamically determined reaction force for the rolling TT, $\mathbf{F}=mg\hat{z}-m\frac{d}{dt}(\boldsymbol{\omega}\times\mathbf{a})$, is not vertical in general, so the condition $\mathbf{F}=g_n\hat{z}$ is a further restriction for rolling solutions of the TT equations.

These solutions play a central role in understanding the inversion of TT because they belong to the asymptotic LaSalle set $\{(\mathbf{L},\mathbf{\hat{3}},\mathbf{\dot{s}}): \dot{E}(\mathbf{L},\mathbf{\hat{3}},\mathbf{\dot{s}})=0\}$, which attracts solutions of the TT equations.

 Under the pure rolling condition the external force, as given by our model, is purely vertical: $\mathbf{F}=g_n\hat{z}$. The system of equations becomes
\begin{gather}
\label{EOM_rTT_3}
m\mathbf{\ddot{r}}=0,\qquad \mathbf{\dot{L}}=R\alpha g_n \mathbf{\hat{3}}\times\hat{z},\qquad \mathbf{\dot{\mathbf{\hat{3}}}}=\frac{1}{I_1}\mathbf{L}\times\mathbf{\hat{3}},
\end{gather}
where the third coordinate of the f\/irst equation is determined by $m\ddot{s}_{\hat{z}}=g_n-mg$. Further reduction using the pure rolling condition $\mathbf{\dot{r}}+(\boldsymbol{\omega}\times\mathbf{a})_{\hat{x},\hat{y}}=0$ restricted to the plane of support yields the autonomous system
\begin{gather}
\label{EOM_rTT_2}
\mathbf{\dot{L}}=R\alpha g_n\mathbf{\hat{3}}\times\hat{z},\qquad
\mathbf{\dot{\mathbf{\hat{3}}}}=\frac{1}{I_1}\mathbf{L}\times\mathbf{\hat{3}},
\end{gather}
along with the constraint $\frac{d}{dt}(\boldsymbol{\omega}\times\mathbf{a})_{\hat{x},\hat{y}}=0$ (the subscripts $\hat{x}$, $\hat{y}$ indicate that the vector is restricted to the supporting plane). But as shown in~\cite{Eben} and~\cite{RSG}, we have:
\begin{proposition}\label{constr_prop}
For the solutions to \eqref{EOM_rTT_2} the constraint $\frac{d}{dt}(\boldsymbol{\omega}\times\mathbf{a})_{\hat{x},\hat{y}}=0$ can be written as
\begin{gather*}
\hat{z}\cdot(\mathbf{L}\times\mathbf{\hat{3}})=0,\qquad \boldsymbol{\omega}\times\mathbf{a}=0.
\end{gather*}
\end{proposition}

These conditions def\/ine an invariant manifold of solutions to equations \eqref{EOM_rTT_2}.
Thus for the solutions of the rolling TT under the assumptions of our model we have that: the vectors~$\mathbf{L}$,~$\hat{z}$ and~$\mathbf{\hat{3}}$ lie in the same plane and the angular velocity~$\boldsymbol{\omega}$ is parallel to the vector~$\mathbf{a}$. This in turn implies that the~$CM$ remains stationary ($\mathbf{\dot{s}}=0$).
These are the situations where the TT is either spinning in the upright position, spinning in the inverted position or the TT is rolling around the $\mathbf{\hat{3}}$-axis in such a way that the~$CM$ is f\/ixed (\emph{tumbling solutions}).

 In terms of the Euler angles, \eqref{EOM_rTT_3} gives rise to three equations of motion and two constraints:
\begin{gather}
-I_1\ddot{\varphi}\sin\theta-2I_{1}\dot{\varphi}\dot{\theta}\cos\theta+I_3\omega_3\dot{\theta}=0,\nonumber\\
 I_1\ddot{\theta}+I_3\dot{\varphi}\omega_3\sin\theta-I_1\dot{\varphi}^2\cos\theta\sin\theta=-R\alpha g_n\sin\theta,\nonumber\\
 I_3\dot{\omega}_3=0,\nonumber\\
 mR\big(\ddot{\theta}(1-\alpha\cos\theta)+\alpha\dot{\theta}^2\sin\theta
 +\dot{\varphi}^2\sin\theta(\alpha-\cos\theta)+\dot{\varphi}\omega_3\sin\theta\big)=0,\nonumber\\
 mR\big(\dot{\omega}_3\sin\theta+\omega_3\dot{\theta}\cos\theta
 +\ddot{\varphi}\sin\theta(\alpha-\cos\theta)+2\dot{\varphi}\dot{\theta}\cos\theta(\alpha-\cos\theta)\big)=0,
\label{rTT_euler_1-5}
\end{gather}
and the quantity $g_n$ is determined by~\eqref{g_n_TT_euler} with $\nu_x=0$.

 After substituting the equations of motion into the constraint equations, we end up with the following conditions:
\begin{gather}
\sin\theta\big(\alpha I_1(\dot{\theta}^2+\dot{\varphi}^2\sin^2\theta)+\dot{\varphi}\omega_3(I_1-I_3+\alpha I_3\cos\theta)-R\alpha g_n(1-\alpha\cos\theta)\big)=0,\nonumber\\
\omega_3\dot{\theta}(\alpha I_3+(I_1-I_3)\cos\theta)=0.\label{rTT_constr_1}
\end{gather}
Using these conditions, we can determine admissible types of solutions to~\eqref{EOM_rTT_3}. As shown in~\cite{Nisse}, these constraints imply $\dot{\theta}\sin\theta=0$, or that $\theta$ is constant for these solutions to the rolling TT. Note that this is the condition $\hat{z}\cdot(\mathbf{L}\times\mathbf{\hat{3}})=0$ from Proposition~\ref{constr_prop} when expressed in terms of Euler angles.

 We see that for $\theta=0$ or $\theta=\pi$ we have the upright or inverted spinning TT, and for constant $\theta\in(0,\pi)$ the f\/irst constraint equation~\eqref{rTT_constr_1} gives
\begin{gather}
\label{constr_rTT_1.1}
\alpha I_1\dot{\varphi}^2\sin^2\theta+\dot{\varphi}\omega_3(I_1-I_3+\alpha I_3\cos\theta)-R\alpha g_n(1-\alpha\cos\theta)=0.
\end{gather}
The f\/irst equation of~\eqref{rTT_euler_1-5} implies that $\ddot{\varphi}=0$ (i.e.~$\dot{\varphi}$ is constant) and the second equation of~\eqref{rTT_euler_1-5} gives
\begin{gather}
\label{rTT_euler_2.1}
I_3\dot{\varphi}\omega_3-I_1\dot{\varphi}^2\cos\theta=-R\alpha g_n.
\end{gather}
By eliminating $g_n$ between \eqref{constr_rTT_1.1} and \eqref{rTT_euler_2.1} we obtain $I_{1}\dot{\varphi}^2(\alpha-\cos\theta)+I_1\dot{\varphi}\omega_3=0$. Here $\dot{\varphi}\neq 0$. The opposite would lead to $g_n\sin\theta=0$ in \eqref{rTT_euler_1-5} (thus contradicting the assumption $\sin\theta\neq0$) and so
\begin{gather*}
\omega_3=\dot{\varphi}(\cos\theta-\alpha).
\end{gather*}
Using again \eqref{rTT_euler_2.1} with the expression for $g_n$ we get a second equation for $\dot{\theta}$ and $\omega_3$:
\begin{gather*}
I_3\dot{\varphi}\omega_3-I_1\dot{\varphi}^2\cos\theta= \frac{-mgR\alpha I_1-mR^2\alpha^2\sin^2\theta(I_1\dot{\varphi}^2\cos\theta-I_3\omega_3\dot{\varphi})}{I_1+mR^2\alpha^2\sin^2\theta}\qquad
\Leftrightarrow \nonumber\\
I_3\dot{\varphi}\omega_3-I_1\dot{\varphi}^2\cos\theta= -mR\alpha g.
\end{gather*}
From this relation we easily deduce that $g_n=mg$ for rolling solutions to TT with vertical reaction force (this is obvious if $\sin\theta=0$). We summarize these results in a proposition.
\begin{proposition}
The solutions to the system \eqref{EOM_rTT_2} under the constraint  $\frac{d}{dt}(\boldsymbol{\omega}\times\mathbf{a})_{\hat{x},\hat{y}}=0$ are either spinning solutions $\theta=0,\,\pi$, or tumbling solutions characterized by $\dot{\theta}=0$, $\theta\in(0,\pi)$, where $\dot{\varphi}$ and $\omega_3$ are determined by the system
\begin{gather}
\label{om_phi_rel}
\omega_3+\dot{\varphi}(\alpha-\cos\theta)=0,\qquad I_3\dot{\varphi}\omega_3-I_1\dot{\varphi}^2\cos\theta=-mR\alpha g.
\end{gather}
\end{proposition}
Note that if we go back to the angular velocity $\dot{\psi}$, the f\/irst equation above becomes $\dot{\psi}+\alpha\dot{\varphi}=0$. This is the condition found by Pliskin \cite{Plisk} where it came up by considering a precessing TT with stationary $CM$. These are the tumbling solutions.

 We can formally solve \eqref{om_phi_rel} to obtain equations for the angular velocities as functions of the inclination angle:
\begin{gather}
 \dot{\varphi}= \pm\sqrt{\dfrac{mR\alpha g}{I_3(\alpha-\cos\theta)+I_1\cos\theta}},\qquad
\omega_3= \mp(\alpha-\cos\theta)\sqrt{\dfrac{mR\alpha g}{I_3(\alpha-\cos\theta)+I_1\cos\theta}}. \label{om_phi_equs}
\end{gather}
These equations together with the value of the Jellett integral $\lambda=R(I_1\dot{\varphi}\sin^2\theta-I_3\omega_3(\alpha-\cos\theta))$, determine signs in equations~\eqref{om_phi_equs}. The values of the parameters $\alpha$, $I_1$ and $I_3$ give restrictions whether these equations are def\/ined for all values of $\cos\theta$ in the range $(-1,1)$.

The right hand side of equations \eqref{om_phi_equs} are real for a full range of $\cos\theta$ if $I_3\alpha+(I_1-I_3)\cos\theta>0$ for all $\cos\theta\in(-1,1)$. If $I_1>I_3$ we see by setting $\cos\theta=-1$ that $I_1<I_3(1+\alpha)$. If on the other hand $I_1<I_3$ then by setting $\cos\theta=1$ we have the condition $I_1>I_3(1-\alpha)$. So if the parameters satisfy:
\begin{gather}
\label{gamma_interval}
1-\alpha<\gamma<1+\alpha,
\end{gather}
where $\gamma=I_1/I_3$, then $\dot{\varphi}$ and $\omega_3$ in \eqref{om_phi_equs} are def\/ined and real for all $\cos\theta\in(-1,1)$. For~$\gamma$ outside this interval, $\dot{\varphi}$~and~$\omega_3$ will be real for $\cos\theta$ belonging to a subinterval of $(-1,1)$, which is characterized by the parameters $\gamma$ and~$\alpha$.
Given a value of Jellett's integral~$\lambda$, we can investigate these intervals for when tumbling solutions exist and the number of tumbling trajectories for that value (see~\cite{RSG}).

 In \cite{Eben}, the relative stability (in the sense of Lyapunov) of a given spinning and tumbling solution is derived as a relation between the integral $\lambda$ (specifying initial conditions for the TT in terms of the angular momentum) and physical characteristics of the TT.

Such a solution is stable if, given a value for $\lambda$, $E_{\lambda}(\theta)$ has a minimum, where $E_{\lambda}(\theta)$ is \eqref{MErTT} under the conditions $\dot{\theta}=0$ and $\omega_3+\dot{\varphi}(\alpha-\cos\theta)=0$ (which is equivalent to $\lambda\sqrt{d(\cos\theta)}(\alpha-\cos\theta)+DR(\gamma\sin^2\theta+(\alpha-\cos\theta)^2)=0$):
\begin{gather*}
E_{\lambda}(\theta)= V\left(\cos\theta,D=-\big(\lambda\sqrt{d(\cos\theta)}(\alpha-\cos\theta)\big)/\big(R\big(\gamma\sin^2\theta+(\alpha-\cos\theta)^2\big)\big),\lambda\right)\nonumber\\
\phantom{E_{\lambda}(\theta)}{} = \frac{\lambda^2}{2R^2(I_1(1-\cos^2\theta)+I_3(\alpha-\cos\theta)^2)}+mgR(1-\alpha\cos\theta),
\end{gather*}
which is the form for the potential for steady motions of an axisymmetric sphere, derived in~\mbox{\cite{Karap3, Karap2}}.

In particular, if $|\lambda|$ is above the threshold value $\frac{\sqrt{mgR^3I_3\alpha}(1+\alpha)^2}{\sqrt{1+\alpha-\gamma}}$, only the inverted spinning position $\cos\theta=-1$ is stable~\cite{Eben,Rau}. This analysis thus specif\/ies how fast a given TT should be spun to make solutions to~\eqref{EOM_rTT_3} stable.

 We have characterized the solutions to~\eqref{EOM_TT_constraint} such that $\mathbf{v}_A=0$ and $\mathbf{F}=g_n\hat{z}$. Clearly solutions to~\eqref{EOM_rTT_2} are also solutions to the system~\eqref{EOM_TT_3}. To be precise, they are part of the precessional solutions, for which the angle $\theta$ is constant. The question is whether these are also asymptotic solutions to the system \eqref{EOM_TT_constraint}.

 In \cite{RSG} a theorem of LaSalle type \cite{LaS} has been formulated to conf\/irm this fact. It is shown that each solution to \eqref{EOM_TT_constraint} satisfying the contact criterion $\hat{z}\cdot(\mathbf{a+s})=0$ and such that $g_n(t)\geq 0$ for $t\geq 0$ approaches exactly one solution to \eqref{EOM_rTT_3} as $t\to\infty$. This is because the trajectories drawn by the solutions to~\eqref{EOM_rTT_3} constitute an invariant submanifold to the system \eqref{EOM_TT_constraint} and belong to the largest invariant set in the LaSalle set $\{(\mathbf{L},\mathbf{\hat{3}},\mathbf{\dot{s}}): \dot{E}(\mathbf{L},\mathbf{\hat{3}},\mathbf{\dot{s}})=0\}$. Thus the solution set to~\eqref{EOM_rTT_3} can be seen as an asymptotic set for the solutions of the TT equations.

 Analysis of special solutions of the TT when $\mathbf{v}_A=0$ thus leads to the most important conclusion about TT behaviour. The application of LaSalle's theorem says that trajectories of the system~\eqref{EOM_TT_constraint} approach the asymptotic set, which consists of spinning and tumbling solutions. From the analysis of the equations of motion in the Euler angle form it follows that tumbling solutions exist for all $\cos\theta\in (-1,1)$ if \eqref{gamma_interval} is satisf\/ied. If moreover $\lambda$ is above the threshold value $\frac{\sqrt{mgR^3I_3\alpha}(1+\alpha)^2}{\sqrt{1+\alpha-\gamma}}$, only the inverted position is stable. Thus we have conditions for when an inverted spinning solution is the only attractive asymptotic state in the asymptotic LaSalle set. Therefore a TT satisfying these condition has to invert. But analysis of dynamics of inversion is still in its infancy.

\section{Reductions of TT equations to equations\\ for related rigid bodies}

In this section we present two reductions of the TT equations that, remarkably, describe motion of simpler rigid bodies. They are given by dif\/ferential equations that are simpler but nevertheless retain certain essential features of the TT equations. They are interesting in their own right and may also work as a testing ground for ideas of how to analyse equations of TT.

\subsection{The gliding HST}

The gliding heavy symmetric top (HST) equations represent an axially symmetric top that is allowed to glide in the supporting plane.

\begin{figure}[ht]
\centering
\includegraphics[scale=0.80]{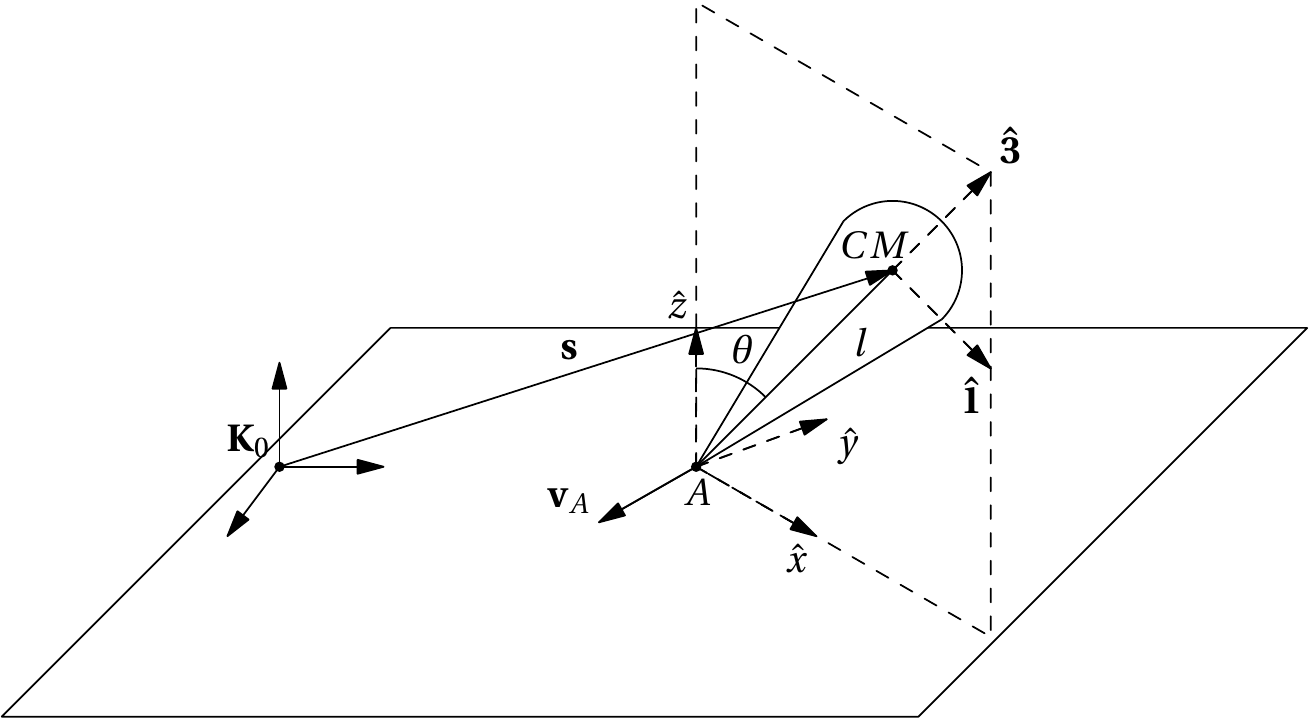}
\caption{Diagram of the gliding HST. Note that $\hat{x}$ is in the plane spanned by $\hat{z}$ and $\mathbf{\hat{3}}$.}\label{glHST_diagram}
\end{figure}

The movements of the gliding HST are described using the same reference systems as before. In Fig.~\ref{glHST_diagram} we have placed $\mathbf{K}=(\hat{x},\hat{y},\hat{z})$ with origin at the contact point~$A$, but this is not essential in our calculations. We let $\mathbf{s}$ be the position vector for the~$CM$ in the inertial system~$\mathbf{K}_0$ and $\mathbf{a}=-l\mathbf{\hat{3}}$ is the vector from~$CM$ to~$A$. We shall assume the contact criterion, i.e.\ that $\hat{z}\cdot(\mathbf{s}(t)+\mathbf{a}(t))\stackrel{t}{\equiv}0$.

 We consider equations of motion for the gliding HST with respect to the $CM$, and the only force creating a torque on $CM$ is the external force $\mathbf{F}$ acting at A and having the moment arm~$\mathbf{a}$. For the force applied at $A$ we have $\mathbf{F}=-\mu g_n\mathbf{v}_A+g_n\hat{z}$. The equations of motion~\eqref{EOM_TT_red1} specialize~to:
\begin{gather}
\label{EOM_gHST_2}
\mathbf{\dot{L}}=(-l\mathbf{\hat{3}})\times\left(g_n\hat{z}-\mu g_n\mathbf{v}_A\right),\qquad m\mathbf{\ddot{r}}=-\mu g_n\mathbf{v}_A,\qquad\mathbf{\dot{\mathbf{\hat{3}}}}=\boldsymbol{\omega}\times\mathbf{\hat{3}},
\end{gather}
This is a system of 11 equations for the variables $(\mathbf{L},\mathbf{r},\mathbf{\dot{r}},\mathbf{\hat{3}})$ and for the value of the normal force~$g_n$. The constraint $0=\hat{z}\cdot(\mathbf{s}+\mathbf{a})$ and its time derivatives $\frac{d}{dt}(\mathbf{s}+\mathbf{a})\cdot\hat{z}=0$, $\frac{d^2}{dt^2}(\mathbf{s}+\mathbf{a})\cdot\hat{z}=0$ determine $s_{\hat{z}}$, $\dot{s}_{\hat{z}}$ and
\begin{gather*}
g_n=\frac{mgI_1-ml(\cos\theta(I_1\dot{\varphi}^2\sin^2\theta+I_1\dot{\theta}^2)
-I_3\dot{\varphi}\omega_3\sin^2\theta)}{I_1+ml^2\sin^2\theta+ml^2\mu\nu_x\cos\theta\sin\theta}.
\end{gather*}
 System \eqref{EOM_gHST_2} rewritten in Euler angles and solved w.r.t.\ $(\ddot{\theta},\ddot{\varphi},\dot{\omega}_3,\dot{\nu}_x,\dot{\nu}_y)$:
\begin{gather}
\label{EOM_gHST_Euler_1}\ddot{\theta}= \frac{1}{I_1}\left(I_1\dot{\varphi}^2\sin\theta\cos\theta-I_3\omega_3\dot{\varphi}\sin\theta+l\mu g_n\nu_x\cos\theta+lg_n\sin\theta\right),\\
\label{EOM_gHST_Euler_2}\ddot{\varphi}= \frac{1}{I_1\sin\theta}\big(I_3\omega_3\dot{\theta}-2I_1\dot{\theta}\dot{\varphi}\cos\theta+l\mu g_n\nu_y\big),\\
\label{EOM_gHST_Euler_3}\dot{\omega}_3= 0,\\
\dot{\nu}_x= \frac{l\sin\theta}{I_1}\big(I_3\omega_3\dot{\varphi}\cos\theta+I_1\big(\dot{\theta}^2+\dot{\varphi}^2\sin^2\theta\big)-lg_n\cos\theta\big)\nonumber\\
\hphantom{\dot{\nu}_x=}{} -\frac{\mu g_n\nu_x}{mI_1}\big(I_1+ml^2\cos^2\theta\big)+\nu_y\dot{\varphi}, \label{EOM_gHST_Euler_4}\\
\label{EOM_gHST_Euler_5}\dot{\nu}_y= -\frac{lI_3\omega_3\dot{\theta}}{I_1}-\frac{I_{1}+ml^2}{mI_1}\mu g_n\nu_y-\nu_x\dot{\varphi}.
\end{gather}
We see that $L_{\mathbf{\hat{3}}}=I_3\omega_3$ is an integral of motion, due to rotational symmetry about the $\mathbf{\hat{3}}$-axis. The HST with f\/ixed supporting point $A$ also admits \[
L_{\hat{z}}=\mathbf{L}_{A}\cdot\hat{z}=\left(\mathbf{L}
+m\mathbf{a}\times(\boldsymbol{\omega}\times\mathbf{a})\right)\cdot\hat{z}=I_{1}^*\dot{\varphi}\sin^2\theta+I_3\omega_3\cos\theta
 \]
 as an integral of motion (where $I_{1}^*=I_1+ml^2$ is the moment of inertia w.r.t.~$A$). For the gliding HST however, this quantity is not an integral of motion since:
\begin{gather*}
\dot{L}_{\hat{z}}= \mathbf{\dot{L}}_{A}\cdot\hat{z}=\frac{d}{dt}\left(\mathbf{L}+m\mathbf{a}\times(\boldsymbol{\omega}\times\mathbf{a})\right)\cdot\hat{z}
=(\mathbf{a\times F})\cdot\hat{z}+m\left(\mathbf{a}\times\left(\frac{d}{dt}(\boldsymbol{\omega}\times\mathbf{a})\right)\right)\cdot\hat{z}\\
\hphantom{\dot{L}_{\hat{z}}}{} = (\mathbf{a}\times(m\mathbf{\ddot{s}}+mg\hat{z}))\cdot\hat{z}+m\mathbf{a}\times\left(\frac{d}{dt}(\boldsymbol{\omega}\times\mathbf{a})\right)
=m(\mathbf{a}\times\mathbf{\dot{v}}_A)\cdot\hat{z}\\
\hphantom{\dot{L}_{\hat{z}}}{} = \frac{I_3\sin\theta}{I_1}ml^2\omega_3\dot{\theta}+\frac{I_{1}^*\sin\theta}{I_1}l\mu g_n\nu_y
\end{gather*}
(according to \eqref{EOM_gHST_Euler_4} and \eqref{EOM_gHST_Euler_5}).

 Due to the presence of the frictional force, the gliding HST is not a conservative system. So the dif\/ferentiation of the energy function $E=\frac{1}{2}m\mathbf{\dot{s}}^2+\frac{1}{2}\boldsymbol{\omega}\cdot\mathbf{L}+mg\mathbf{s}\cdot\hat{z}$ gives the same result as for the TT: $\dot{E}=\mathbf{F}\cdot\mathbf{v}_A$. The derivative of the modif\/ied energy function, i.e.\ the part of the energy not involving $\mathbf{v}_{A}$, satisf\/ies $\frac{d}{dt}\tilde{E}(t)=m\mathbf{\dot{v}}_{A}\cdot(\boldsymbol{\omega}\times\mathbf{a})$, as for TT (see~\eqref{diff_Et}).

 With the properties of the three functions $L_\mathbf{\hat{3}}$, $L_{\hat{z}}(t)$ and $\tilde{E}(t)$ established, we see that the equations of motion \eqref{EOM_gHST_Euler_1}--\eqref{EOM_gHST_Euler_5} are equivalent to the system of dif\/ferential equations
\begin{gather*}
 \frac{d}{dt}L_{\mathbf{\hat{3}}}=0,\\
\frac{d}{dt}L_{\hat{z}}(t)=m(\hat{z}\times\mathbf{a})\cdot\mathbf{\dot{v}}_{A}=\frac{I_3\sin\theta}{I_1}ml^2\omega_3\dot{\theta}+\frac{I_{1}^*\sin\theta}{I_{1}}l\mu g_n\nu_y,\\
\frac{d}{dt}\tilde{E}(t)=m(\boldsymbol{\omega}\times\mathbf{a})\cdot\mathbf{\dot{v}}_A
=-ml((\dot{\nu}_x-\dot{\varphi}\nu_y)\dot{\theta}\cos\theta+\dot{\varphi}\sin\theta(\dot{\nu}_y+\dot{\varphi}\nu_x)),\\
\frac{d}{dt}m\mathbf{\dot{r}}=-\mu g_n\mathbf{v}_A,
\end{gather*}
in the same way as the system \eqref{EOM_TT_diff} is equivalent to equations \eqref{EOM_TT_Euler_1}--\eqref{EOM_TT_Euler_5}. If we substitute the expressions for $\omega_3=\frac{L_{\mathbf{\hat{3}}}}{I_3}$, $\dot{\varphi}=\frac{L_{\hat{z}}(t)-L_\mathbf{\hat{3}}\cos\theta}{I_{1}^*\sin^2\theta}$, into the expression for the modif\/ied energy function we get, analogously to~\eqref{METT}, the Main Equation for the gliding HST (MEgHST):
\begin{gather}
\label{MEgHST}
\tilde{E}(t)=\frac{I_{1}^*\dot{\theta}^2}{2}+\frac{L_{\mathbf{\hat{3}}}^2}{2I_3}+\frac{(L_{\hat{z}}(t)-L_{\mathbf{\hat{3}}}\cos\theta )^2}{2I_{1}^*\sin^2\theta}+mgl\cos\theta=\frac{1}{2}I_{1}^*\dot{\theta}^2+V(\cos\theta,L_\mathbf{\hat{3}},L_{\hat{z}}).
\end{gather}

\subsection{Transformation from TT to gliding HST}

The gliding HST equations have been introduced here because they appear as a limit of the TT equations when $R\alpha=-l$ and $R\to 0$. This also transforms integrals of TT into the integrals of the gliding HST as well as METT~\eqref{METT} into MEgHST~\eqref{MEgHST}.

\begin{figure}[ht]
\centering
\includegraphics[scale=0.80]{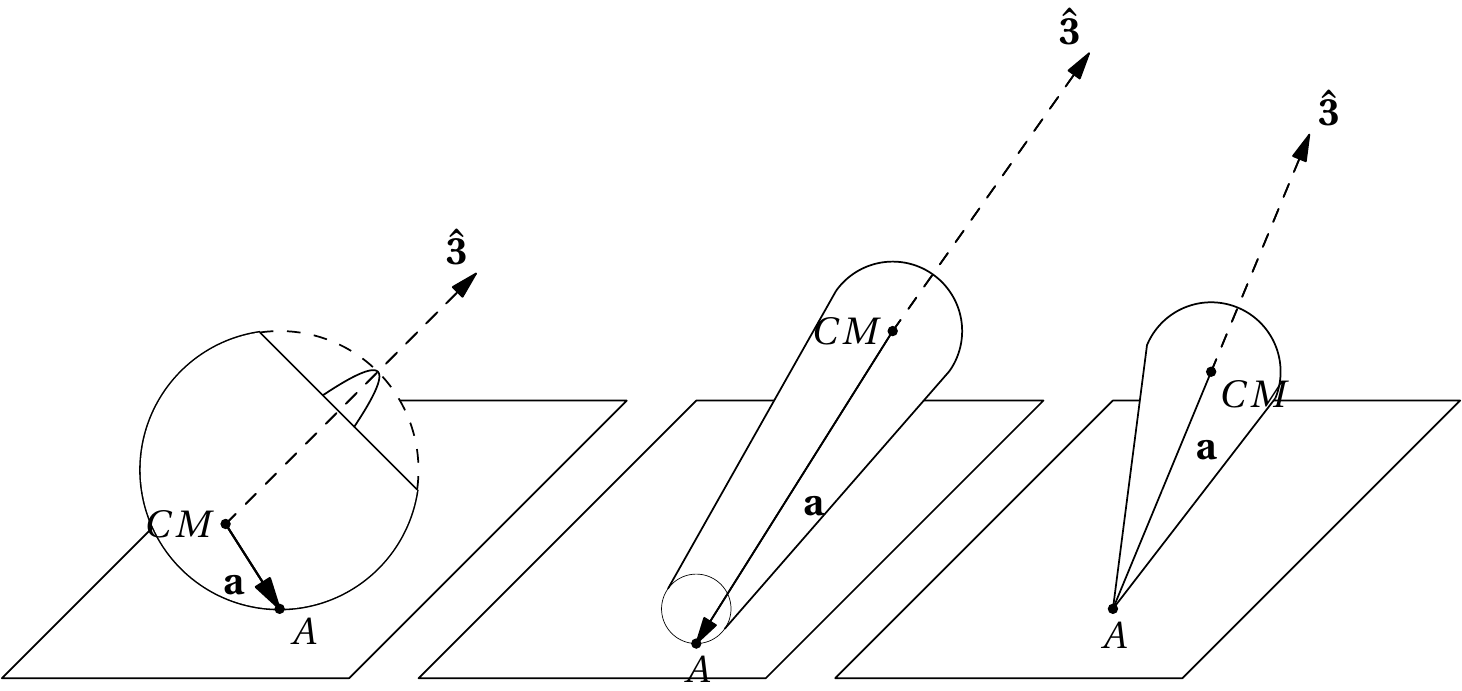}
\caption{Transformation from TT to gliding HST: $R\alpha=-l$, $R\to0$.\label{transform_TT}}
\end{figure}

 The vector $\mathbf{a}$ connecting the center of mass to the point of contact is $R(\alpha\mathbf{\hat{3}}-\hat{z})$ for the TT system, and is $-l\mathbf{\hat{3}}$ for the gliding HST. To describe the transformation from TT to gliding HST we think of the body of TT being stretched (still maintaining its axial symmetry and the spherical bottom), and the spherical part being shrunk to a point. So it looks like a ball-point pen during this transformation. The center of mass thus moves up along the $\mathbf{\hat{3}}$-axis and the radius of the sphere becomes small. We take the limit $R\to0$ subjected to the condition $R\alpha=-l$ (see Fig.~\ref{transform_TT}).
 For the vector equations, the thing that changes is the vector $\mathbf{a}$. We directly see that the equations of motion for the TT,
\begin{gather*}
m\mathbf{\ddot{s}}=\mathbf{F}-mg\hat{z},\qquad \mathbf{\dot{L}}=R(\alpha\mathbf{\hat{3}}-\hat{z})\times\mathbf{F},\qquad \mathbf{\dot{\mathbf{\hat{3}}}}=\boldsymbol{\omega}\times\mathbf{\hat{3}},
\end{gather*}
become the equations of motion for the gliding HST:
\begin{gather*}
m\mathbf{\ddot{s}}=\mathbf{F}-mg\hat{z},\qquad \mathbf{\dot{L}}=-l\mathbf{\hat{3}}\times\mathbf{F},\qquad \mathbf{\dot{\mathbf{\hat{3}}}}=\boldsymbol{\omega}\times\mathbf{\hat{3}},
\end{gather*}
since $\mathbf{a}=R\alpha\mathbf{\hat{3}}-R\hat{z}=-l\mathbf{\hat{3}}-R\hat{z}\to-l\mathbf{\hat{3}}$ as $R\to 0$.

 The dif\/ference between these dynamical systems is that the TT is allowed to glide and roll in the plane, whereas the gliding HST glides and rotates. In both cases the body in question satisf\/ies the contact constraint.
We can also see the ef\/fects of the transformation applied to the equations of motion for TT  \eqref{ddth}--\eqref{nu_y} expressed in the Euler angles. If we substitute $\alpha R$ with~$-l$ everywhere in the equations and then let $R\to 0$, we arrive at equations~\eqref{EOM_gHST_Euler_1}--\eqref{EOM_gHST_Euler_5}.

 The limiting form of Jellett's integral $\lambda$ and of Routh's function is more tricky and does not immediately
 lead to $L_\mathbf{\hat{3}}$ and $L_{\hat{z}}$ for the gliding HST. This is not surprising since~$\lambda$ and $D$ are derived for a spherical body.

 For the Jellett integral we get
 \[
 \lambda=lI_{3}\omega_3+R(I_1\dot{\varphi}\sin^2\theta+I_3\omega_3\cos\theta)= lL_{\mathbf{\hat{3}}}+O(R)\to lL_{\mathbf{\hat{3}}}.
 \]
This transformation is also obvious in vector notation, since $\lambda=-\mathbf{L}\cdot\mathbf{a}=R(L_{\hat{z}}-\alpha L_{\mathbf{\hat{3}}})\to lL_\mathbf{\hat{3}}$ as $R\alpha=-l$, $R\to0$. For the Routh function we expand with respect to small~$R$:
\begin{gather*}
D= I_3\omega_3\sqrt{d(\cos\theta)}=I_3\omega_3\sqrt{\frac{I_1}{I_3}+\frac{ml^2}{I_3}
+\frac{2mlR}{I_3}\cos\theta+\frac{mR^2}{I_3}\left(\frac{I_1}{I_3}\sin^2\theta+\cos^2\theta\right)}\nonumber\\
\phantom{D}{} = \omega_3\sqrt{I_3(I_1+ml^2)}\left[1+R\frac{2ml\cos\theta}{I_1+ml^2}+\frac{mR^2}{I_1+ml^2}\big(I_1\sin^2\theta+I_3\cos^2\theta\big)\right]^{1/2}\nonumber\\
\phantom{D}{} = \omega_3\sqrt{I_3I_{1}^*}\left(1+R\frac{ml\cos\theta}{I_{1}^*}+O\big(R^2\big)\right).
\end{gather*}
Combining this with $\lambda$,
\begin{gather*}
\frac{I_{1}^{*}\lambda-l\sqrt{I_3 I_{1}^{*}}D}{I_1 R}= \frac{1}{I_1 R}\bigg[lI_{1}^{*}I_{3}\omega_3+RI_{1}^{*}(I_1\dot{\varphi}\sin^2\theta+I_3\omega_3\cos\theta)\nonumber\\
\phantom{\frac{I_{1}^{*}\lambda-l\sqrt{I_3 I_{1}^{*}}D}{I_1 R}=}{}
-l\omega_3 I_3I_{1}^*\left(1+R\frac{ml\cos\theta}{I_{1}^*}+O\big(R^2\big)\right)\bigg]\nonumber\\
\phantom{\frac{I_{1}^{*}\lambda-l\sqrt{I_3 I_{1}^{*}}D}{I_1 R}}{}  = \frac{1}{I_{1}R}\bigg[R(I_{1}^{*}I_1\dot{\varphi}\sin^2\theta+\left(I_{1}^*-ml^2\right)I_{3}\omega_3\cos\theta)+O\big(R^2\big)\bigg]\nonumber\\
\phantom{\frac{I_{1}^{*}\lambda-l\sqrt{I_3 I_{1}^{*}}D}{I_1 R}}{}
= \left(I_{1}^*\dot{\varphi}\sin^2\theta+I_{3}\omega_3\cos\theta\right)+O(R)=L_{\hat{z}}+O(R),
\end{gather*}
we see that it tends to $L_{\hat{z}}$ as $R\to 0$. We have thus retrieved from $\lambda$ and $D$ both functions $L_{\mathbf{\hat{3}}}$ and $L_{\hat{z}}$.

{\sloppy Slightly more elaborate is the calculation that the METT \eqref{METT} is transformed to MEgHST~\eqref{MEgHST}. It requires careful expansion of the ef\/fective potential $V(\cos\theta,D(t),\lambda)$ with respect to the small variable $R$ (see~\cite{Nisse}).

 }

\begin{proposition}
The METT equation $E(t)=g(\cos\theta)\dot{\theta}^2+V(\cos\theta,D(t),\lambda)$ becomes, under the transformation $R\alpha=-l$, $R\to 0$, the MEgHST:
\begin{gather*}
\tilde{E}(t)=\frac{I_{1}^*\dot{\theta}^2}{2}+\frac{L_{\mathbf{\hat{3}}}^2}{2I_3}
+\frac{(L_{\hat{z}}(t)-L_{\mathbf{\hat{3}}}\cos\theta)^2}{2I_{1}^*\sin^2\theta}+mgl\cos\theta.
\end{gather*}
\end{proposition}
 The MEgHST has simpler ef\/fective potential
 \[
 V(\cos\theta,L_{\hat{z}}(t),L_\mathbf{\hat{3}})
 =\frac{L_{\mathbf{\hat{3}}}^2}{2I_3}+\frac{(L_{\hat{z}}(t)-L_{\mathbf{\hat{3}}}\cos\theta)^2}{2I_{1}^*\sin^2\theta}+mgl\cos\theta
\]
  than METT, but retains the same structure.

\subsection{The gliding eccentric cylinder}

It is also possible to reduce the TT equations so they describe an eccentric rolling and gliding cylinder (glC). These equations are a special case of the TT equations when $\lambda=D=0$.

\begin{figure}[ht]
\centering
\includegraphics[scale=0.90]{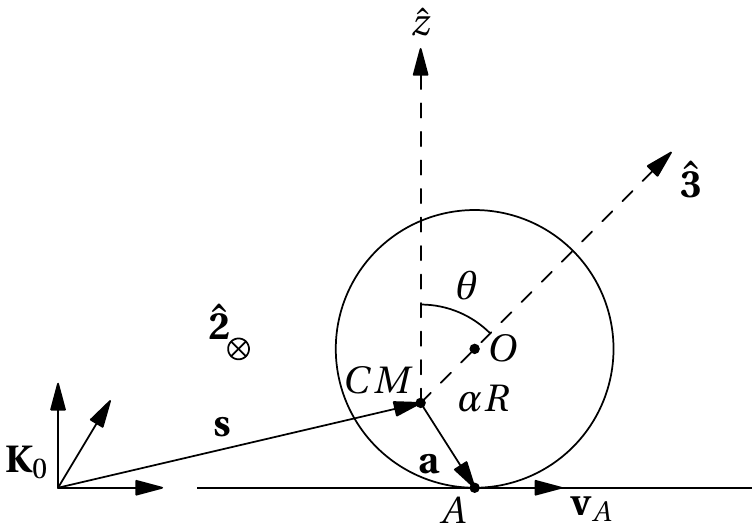}
\caption{Diagram of an eccentric cylinder.\label{cyl_diagram}}
\end{figure}

 We think of an eccentric cylinder, homogeneous in the $\hat{y}$-direction, rolling and gliding on a~supporting plane in the direction of the $\hat{x}$-axis (the cylinder is considered to be long). We could also see it as a model for an eccentric wheel rolling along the $\hat{x}$-axis in the $(\hat{x},\hat{y})$-plane. It has mass $m$ and radius~$R$. The center of mass $CM$ is shifted with respect to the symmetry axis of the cylinder by $\alpha R$, $0<\alpha<1$. We let the system $\mathbf{\tilde{K}}=(\mathbf{\hat{1}},\mathbf{\hat{2}},\mathbf{\hat{3}})$ have its origin in $CM$, where the $\mathbf{\hat{3}}$-axis points along the line determined by $CM$ and the center of the cross-section circle in the direction from $CM$ toward this center. $\mathbf{\hat{1}}$ is orthogonal to $\mathbf{\hat{3}}$ and $\mathbf{\hat{2}}=\mathbf{\hat{3}}\times\mathbf{\hat{1}}$ is parallel to the $\hat{y}$-axis (see Fig.~\ref{cyl_diagram}).

 We let $\theta$ be the angle between the $\hat{z}$-axis and the $\mathbf{\hat{3}}$-axis. This will be the single Euler angle since the cylinder is only rolling in one direction, so that $\boldsymbol{\omega}=\dot{\theta}\mathbf{\hat{2}}$. Further, we let $A$ denote the point of contact of the cylinder with the supporting line, so that the vector from $CM$ to $A$ is $\mathbf{a}=R(\alpha\mathbf{\hat{3}}-\hat{z})$ as for the TT.

 As we have done previously, we consider the external force acting on the cylinder to be $\mathbf{F}=g_n\hat{z}-\mu g_n\mathbf{v}_A$, where $\mathbf{v}_A=\nu_x\hat{x}$ is the velocity of the point $A$. The equations of motion for the glC have the familiar form: $m\mathbf{\ddot{s}}=\mathbf{F}-mg\hat{z}$ and $\mathbf{\dot{L}}=\mathbf{a}\times\mathbf{F}.$

Observe that there is no third equation (for $\mathbf{\hat{3}}$) since the axis $\mathbf{\hat{2}}$ has constant direction parallel to $\hat{y}$.
As the contact criterion $\hat{z}\cdot(\mathbf{s+a})\stackrel{t}{\equiv}0$ determines the vertical component of $\mathbf{s}$, we reduce these equations to
\begin{gather}
\label{EOM_cyl}
 m\mathbf{\ddot{r}}=-\mu g_n\mathbf{v}_A,\qquad \mathbf{\dot{L}}=\mathbf{a}\times\mathbf{F},
\end{gather}
where $\mathbf{r}=\mathbf{s}-s_{\hat{z}}\hat{z}$ is the planar component of the position vector for the $CM$.

This is a system of only two equations: one for $\theta$ and one for $\nu_x$. We thus proceed directly to the coordinate form of the equations of motion. Note that $\mathbf{L}=\mathbb{I}\boldsymbol{\omega}=I\dot{\theta}\mathbf{\hat{2}}$, where $I=I_2$ is the moment of inertia with respect to the axis $\mathbf{\hat{2}}$ going through $CM$. We get
\begin{gather}
 I\ddot{\theta}\mathbf{\hat{2}}=R\mu g_n\nu_x(1-\alpha\cos\theta)\mathbf{\hat{2}}-R\alpha g_n\sin\theta\mathbf{\hat{2}},\nonumber\\
 m\dot{\nu}_x\hat{x}+mR\ddot{\theta}(1-\alpha\cos\theta)\hat{x}+mR\alpha\dot{\theta}^2\sin\theta\hat{x}=-\mu g_n\nu_x\hat{x}. \label{EOM_euler_cyl_1}
\end{gather}
The equations of motion in the form solved w.r.t.\ highest derivative are then
\begin{gather*}
\ddot{\theta} =\frac{R\mu g_n\nu_x}{I}(1-\alpha\cos\theta)-\frac{R\alpha g_n}{I}\sin\theta,\\
\dot{\nu}_x=\frac{R^2\alpha g_n}{I}\sin\theta(1-\alpha\cos\theta)-\frac{\mu g_n\nu_x}{mI}\left(I+mR^2(1-\alpha\cos\theta)^2\right)-R\alpha\dot{\theta}^2\sin\theta.
\end{gather*}
The function $g_n$ is determined in the same way as \eqref{g_n_TT_euler}:
\begin{gather*}
g_n=\frac{mgI+mR\alpha I\dot{\theta}^2\cos\theta}{I+mR^2\alpha^2\sin^2\theta-mR^2\alpha\sin\theta(1-\alpha\cos\theta)\mu\nu_x}.
\end{gather*}
 We directly see that we can get these equations from the equations of motion for TT \eqref{ddth}--\eqref{nu_y} in the special case $\dot{\varphi}=\omega_3=\nu_y=0$.This corresponds to $\lambda=0$, $D=0$ for this system, and the constraint
 $\hat{y}\cdot\mathbf{\dot{s}}=0$ is satisf\/ied (since the cylinder does not move in the $\hat{y}$-direction).
 These constraints are consistent with the equations \eqref{ddth}--\eqref{nu_y}.

 When we consider the gliding eccentric cylinder, we have three natural special cases that can be integrated by quadratures:
\begin{itemize}\itemsep=0pt
\item[a)] the noneccentric case $\alpha=0$, $\mathbf{v}_A\neq 0$;
\item[b)] the nongliding case $\mathbf{v}_A=0$, $\alpha\neq 0$;
\item[c)] the nongliding noneccentric case $\alpha=0$ and $\mathbf{v}_A=0$.
\end{itemize}
For these cases we examine the limits $\mathbf{v}_A\to 0$ and/or $\alpha\to 0$ in this model and notice that the limit $\mathbf{v}_A\to0$ leads only to the static solution $\theta(t)={\rm const}$.

 We consider f\/irst the case of the axially symmetric cylinder, $\alpha=0$. The center of mass and the geometric center of the cylinder coincides and the vector from $CM$ to $A$ reduces to $\mathbf{a}=-R\hat{z}$. We use this in~\eqref{EOM_cyl}, so the equations of motion become $I\ddot{\theta}=R\mu g_n\nu_x$, $m\dot{\nu}_x+mR\ddot{\theta}=-\mu g_n\nu_x$, and in the solved form: $\ddot{\theta}=\frac{R}{I}\mu g_n\nu_x$, $\dot{\nu}_x=-\frac{I+mR^2}{mI}\mu g_n\nu_x$.
These two equations imply that $mR\dot{\nu}_x+(I+mR^2)\ddot{\theta}=0$, i.e.\ $mR\nu_x+(I+mR^2)\dot{\theta}$ is constant. We can also obtain these equations by letting $\alpha\to 0$ in~\eqref{EOM_euler_cyl_1}. If we assume that $\mu$ is constant (note here that $g_n=mg$), the system can be solved explicitly for $\nu_x(t)$ and $\dot{\theta}(t)$:
\begin{gather*}
\nu_x(t)=\nu_{x}(0)e^{-\frac{I+mR^2}{I}\mu g t},\\
\dot{\theta}(t)=-\frac{mR}{I+mR^2}\nu_{x}(0)e^{-\frac{I+mR^2}{I}\mu g t}+\frac{1}{I+mR^2}\big(mR\nu_x(0)+\big(I+mR^2\big)\dot{\theta}(0)\big).
\end{gather*}
If we now let $\mathbf{v}_A\to 0$ in these equations we get the single equation $\ddot{\theta}=0$, saying that the cylinder is turning with constant speed. This describes the motion of a non-eccentric cylinder rolling without friction.

 In the second case we consider equations for a rolling eccentric cylinder. This means we let $\mathbf{v}_A=0$ in the equations of motion:
\begin{gather*}
m\mathbf{\ddot{s}}=\mathbf{F}-mg\hat{z},\qquad \mathbf{\dot{L}}=\mathbf{a}\times\mathbf{F}.
\end{gather*}
We eliminate $\mathbf{F}$ in the second equation to get
\[
\frac{d}{dt}\left(\mathbb{I}\boldsymbol{\omega}\right)=\mathbf{a}\times\left(m\mathbf{\ddot{s}}+mg\hat{z}\right)
=\mathbf{a}\times\left(-m\frac{d}{dt}(\boldsymbol{\omega}\times\mathbf{a})+mg\hat{z}\right).
 \]
 So for the Euler angle $\theta$ we get the nonlinear equation:
 \[
 \ddot{\theta}=\frac{-mg\alpha R\sin\theta-mR^2\alpha\dot{\theta}^2\sin\theta}{I+mR^2((\alpha-\cos\theta)^2+\sin^2\theta)},
\]
  but this equation is also found through the energy formula, since
\begin{gather*}
  0=\dot{E}=m\mathbf{\ddot{s}\cdot\dot{s}}+\boldsymbol{\omega}\cdot\mathbf{L}+mg\mathbf{\dot{s}}\cdot\hat{z}\\
  \phantom{0}{}
  = \dot{\theta}\ddot{\theta}\left(I+mR^2(\sin^2\theta+(\alpha-\cos\theta)^2)\right)+\dot{\theta}^3 mR^2\alpha\sin\theta+mgR\alpha\dot{\theta}\sin\theta.
\end{gather*}
So it admits energy as an integral of motion
\[
E=\frac{1}{2}\dot{\theta}^2\big(I+mR^2\big(\sin^2\theta+(\alpha-\cos\theta)^2\big)\big)+mgR(1-\alpha\cos\theta).
\]
 This is a f\/irst order autonomous ODE that can be separated and has a solution expressed by quadratures.
The equation for $\ddot{\theta}$ reduces to the special case of the rolling noneccentric cylinder $\ddot{\theta}=0$ if we let $\alpha\to 0$ in this equation.

 When we take the limit $\mathbf{v}_A\to 0$ in equations \eqref{EOM_euler_cyl_1} we get an additional constraint that $\mathbf{F}=g_n\hat{z}$ is vertical so the result has to dif\/fer from the previous. The system becomes one equation of motion with one constraint:
\begin{gather}
\label{EOM_ecc_roll_cyl}
I\ddot{\theta}=-R\alpha g_n\sin\theta,\\
mR\ddot{\theta}(1-\alpha\cos\theta)+mR\alpha\dot{\theta}^2\sin\theta=0.\nonumber
\end{gather}
The constraint can be rewritten as $\frac{d}{dt} (\dot{\theta}(1-\alpha\cos\theta) )=0$, i.e.\ $\dot{\theta}(1-\alpha\cos\theta)$ is equal to a~constant~$C$.

The value of $g_n$ is deduced from the contact criterion as before: $g_n=\frac{mgI+mR\alpha I\dot{\theta}^2\cos\theta}{I+mR^2\alpha^2\sin^2\theta}$. Using the constraint $\dot{\theta}=\frac{C}{1-\alpha\cos\theta}$ and the expression for $g_n$ in equation~\eqref{EOM_ecc_roll_cyl} we get a polynomial equation in $\cos\theta$ and $\sin\theta$:
\begin{gather*}
\sin\theta\big(IC^2+mR^2C^2\alpha(\alpha-\cos\theta)-mgR(1-\alpha\cos\theta)^3\big)=0.
\end{gather*}
Thus with the external force $\mathbf{F}=g_n\hat{z}-\mu g_n\mathbf{v}_A$ we get only solutions with constant $\theta$ when we let $\nu_x\to 0$ in the equations of motion~\eqref{EOM_euler_cyl_1}.

\section{Conclusions}

In this paper we have discussed several dif\/ferent forms of the TT equations: the vector form, the Euler angle form, the form leading to METT and the form suitable for discussing the gyroscopic balance condition. Each form gives dif\/ferent insight into the character of motion of TT and helps to unveil complicated features of TT dynamics.

 The vector form helps to explain in simple terms why the gliding friction is necessary for inversion of TT. It also is the most suitable for asymptotic analysis of TT equations and leads to formulation of conditions for when the inverted spinning state is the only admissible stable solution. The conditions for the physical parameters of TT and for the initial conditions seem to agree remarkably well with the experimentally observed features of TT motion.

 Analysis of the Euler angle form of TT equations conf\/irms the picture obtained from the vector equations. It also shows that the functions $(\lambda,D,\tilde{E})$, being integrals of motion of the rolling TT, play a special role in understanding the motion of rolling and gliding TT. By eliminating $\dot{\varphi}$ and $\omega_3$ from the energy $\tilde{E}$ we obtain the METT equation having similar form as the separation equation for the rolling TT. From this we can see that during inversion the symmetry axis is performing fast nutational motion within a nutational band that is moving from the north pole to the south pole of the unit sphere $S^2$.

 Since the Euler equations of TT constitute a nonlinear dynamical system of 6 degrees of freedom we have studied ways of simplifying these equations by taking dynamically invariant constraints and certain limits of the physical parameters. Remarkably we have found two such simplif\/ications that even have mechanical interpretation. The limit $\alpha R=-l$, $R\to 0$ leads to the gliding HST equations, which is an interesting system in itself, but the equations are also interesting in the context of the TT equations because they retain all main features of TT but are simpler for analytical treatment.
We have shown also the constraints $\mathbf{\dot{s}}\cdot\hat{y}=0$, $\lambda=0$ and $D=0$ reduces the TT equations to equations for a rolling and gliding eccentric cylinder.

\pdfbookmark[1]{References}{ref}
\LastPageEnding

\end{document}